\documentclass[aps,reprint,superscriptaddress,nofootinbib]{revtex4-2}

\pdfoutput=1

\usepackage{mathtools}
\usepackage{amssymb}
\usepackage{mathrsfs}
\usepackage{bbm}

\usepackage[dvipsnames]{xcolor}
\usepackage[colorlinks, citecolor={blue!70!black}, urlcolor={blue!70!black}, linkcolor={red!70!black},hyperindex,breaklinks]{hyperref}

\newcommand{\Op}{\mathscr{O}}  
\newcommand{\can}{{\rm can}}  
\newcommand{\GC}{{\rm GC}}  
\newcommand{\NATS}{{\rm NATS}}
\newcommand{\ent}{S_{\rm th}}
\newcommand{\hess}{{\rm H}}  
\newcommand{\diag}{{\rm diag}}  
\def\const{ {\rm const.} }   
\def\id{\mathbbm{1}}   
\newcommand{\Tr}{{\rm Tr}}   
\newcommand{\Sites}{N}  
\newcommand{\var}{{\rm var}}  
\newcommand{\Min}{ {\rm min} }   
\newcommand{\Max}{ {\rm max} }  
\newcommand{\0}{ {(0)} }
\newcommand{\1}{ {(1)} }
\newcommand{\2}{ {(2)} }

\newcommand{\KParen}{ {(k)} }

\renewcommand\th{ {\rm th} }

\newcommand*{\bra}[1]{\langle #1\rvert}
\newcommand*{\ket}[1]{\lvert #1 \rangle}
\newcommand*{\braket}[2]{\langle #1 \lvert #2 \rangle}

\newcommand*{\expval}[1]{\left\langle  #1  \right\rangle}

\begin{document}
 
\title{Non-Abelian eigenstate thermalization hypothesis}
\author{Chaitanya~Murthy}
\email{crmurthy@stanford.edu}
\affiliation{Department of Physics, Stanford University, Stanford, CA 94305, USA}
\author{Arman~Babakhani}
\affiliation{Department of Physics, University of Southern California, Los Angeles, CA, 90089, USA}
\affiliation{Information Sciences Institute, Marina Del Rey, CA, 90292, USA}
\author{Fernando~Iniguez}
\affiliation{Department of Physics, University of California, Santa Barbara, CA 93106, USA}
\author{Mark~Srednicki}
\affiliation{Department of Physics, University of California, Santa Barbara, CA 93106, USA}
\author{Nicole~Yunger~Halpern}
\email{nicoleyh@umd.edu}
\affiliation{Joint Center for Quantum Information and Computer Science, NIST and University of Maryland, College Park, MD 20742, USA}
\affiliation{Institute for Physical Science and Technology, University of Maryland, College Park, MD 20742, USA}
\date{\today}

%
%
\begin{abstract}
The eigenstate thermalization hypothesis (ETH) explains why nonintegrable quantum many-body systems thermalize internally if the Hamiltonian lacks symmetries. If the Hamiltonian conserves one quantity (``charge''), the ETH implies thermalization within a charge sector---in a microcanonical subspace. But quantum systems can have charges that fail to commute with each other and so share no eigenbasis; microcanonical subspaces may not exist. Furthermore, the Hamiltonian will have degeneracies, so the ETH need not imply thermalization. We adapt the ETH to noncommuting charges by positing a \emph{non-Abelian ETH} and invoking the \emph{approximate microcanonical subspace} introduced in quantum thermodynamics. Illustrating with SU(2) symmetry, we apply the non-Abelian ETH in calculating local operators' time-averaged and thermal expectation values. In many cases, we prove, the time average thermalizes. However, we find cases in which, under a physically reasonable assumption, the time average converges to the thermal average unusually slowly as a function of the global-system size. This work extends the ETH, a cornerstone of many-body physics, to noncommuting charges, recently a subject of intense activity in quantum thermodynamics.
\end{abstract}

\maketitle

%

%
%
Nonintegrable closed quantum many-body systems thermalize internally, in the absence of conserved observables, or \emph{charges}. 
Few-body operators $\Op$ equilibrate to the expectation values they would have in the canonical state $\rho_\can \propto e^{-\beta H}$.
$H$ denotes the Hamiltonian, whose expectation value 
determines the inverse temperature $\beta$~\cite{DAlessio_16_From}.
The eigenstate thermalization hypothesis (ETH) explains this thermalization~\cite{Deutsch_91_Quantum, Srednicki_94_Chaos, Rigol_08_Thermalization}:
Let $\ket{\alpha}$ denote the energy eigenstates; 
$E_{\alpha}$, the eigenenergies;
and $\Op_{\alpha \alpha'} \coloneqq \bra{\alpha} \Op \ket{\alpha'}$,
matrix elements representing the operator.
$\Op$ and $H$ satisfy the ETH if 
$\Op_{\alpha \alpha'}$ has a certain structure, reviewed below.
If $\Op_{\alpha \alpha'}$ does and $H$ is nondegenerate,
$\Op$ thermalizes: Its time-averaged expectation value 
approximately equals its thermal expectation value.
The difference is of $O( \Sites^{-1} )$, if
$\Sites$ denotes the global system size.
(We use big-$O$ notation as in many-body physics, 
meaning ``scales as.'') 
These results explain behaviors observed numerically and experimentally across condensed matter; atomic, molecular, and optical physics; and high-energy physics~\cite{DAlessio_16_From, Deutsch_18_Review, Chandran_16_AnyonETH, Jansen_19_HolsteinETH, Lu_19_Renyi, Murthy_19_Structure, Langen_15_Ultracold, Lashkari_18_Eigenstate, Bao_19_Eigenstate, Murthy_19_Bounds, Mueller_21_Thermalization}.

The argument for thermalization relies on
the Hamiltonian's nondegeneracy and 
on matrix-element structure.
Both postulates are questionable if $H$ conserves charges~\footnote{
%
We focus on continuous symmetries to bridge ETH studies with 
the emerging subfield of 
the quantum thermodynamics of noncommuting charges (Hermitian operators that generate continuous symmetries).
}.
If $H$ has an Abelian symmetry, 
the energy spectrum can lack degeneracies. 
Since the charges commute, 
they share eigenspaces---charge sectors. 
In each shared sector, the ETH applies.
For example, consider $\Sites$ qubits (quantum two-level systems, or spins).
$H$ can conserve the total spin's $z$-component, $S_z$, 
by being U(1)-symmetric.
The ETH is often applied in an $S_z$ sector,
wherein the ETH holds and implies thermalization.

A non-Abelian symmetry can eliminate 
our recourse to charge sectors:
Such a symmetry is generated by charges that fail to commute with each other and so cannot necessarily have definite values simultaneously---cannot necessarily share sectors governable by the ETH.
Moreover, non-Abelian symmetries force degeneracies on $H$, having multidimensional irreducible representations.
Finally, how $\Op$ transforms under the symmetry operations constrains
the matrix elements $\Op_{\alpha \alpha'}$ in opposition to the ETH.

For example, consider again an $\Sites$-qubit system.
$H$ can conserve the total spin components $S_{a {=} x,y,z}$, 
by being SU(2)-symmetric.
The energy spectrum splits into degenerate multiplets labeled by
total spin quantum numbers $s_\alpha$. 
Only the singlets, whose $s_\alpha = 0$, are simultaneous eigenspaces of $S_{x,y,z}$.
Furthermore, the matrix elements $\Op_{\alpha \alpha'}$ obey the Wigner--Eckart theorem~\cite{Shankar_Quantum_Book},
conflicting with the ETH.

Non-Abelian symmetries are ubiquitous in 
quantum many-body physics~\cite{yang1996symmetry,gilmore_12_lie}.
They grace systems including
complex nuclei and atoms~\cite{wigner1959group},
Heisenberg models in condensed matter~\cite{auerbach1998interacting,Joel_13_Introduction}, gauge theories~\cite{Peskin_95_Introduction}, and Wess-Zumino-Witten models~\cite{Wess_71_Consequences,Witten_83_Global,Witten_84_Non,Novikov_82_Hamiltonian}.
Hence the apparent conflict between non-Abelian symmetries and the ETH impacts our basic understanding of diverse, prominent models.

To overcome the conflict, we propose a \emph{non-Abelian ETH}.
We apply it to SU(2) symmetry 
for simplicity, expecting results to generalize.
Using the non-Abelian ETH, we compute 
two averages of few-body operators $\Op$:
time-averaged and thermal expectation values.
For many operators and initial states, 
the time average agrees with the thermal average:
Differences are $O(\Sites^{-1})$, as without 
noncommuting charges~\cite{Srednicki_96_Thermal, Srednicki_99_Approach}.
For certain operators and initial states, however,
the time average may deviate from the thermal prediction
by anomalously large corrections $\sim \Sites^{-1/2}$.
This result holds under a physically reasonable assumption
about the non-Abelian analog of $\Op_{\alpha \alpha'}$.

Below, we review the conventional ETH.
We then introduce our setup, present the non-Abelian ETH [Eq.~(\ref{eq_nonAbelian_ETH})],
and apply it to calculate operators' thermal and 
time-averaged expectation values.
Finally, we describe opportunities established by our results.
This work extends the ETH, a mainstay of many-body physics, to the more fully quantum domain of noncommuting charges and so to a growing subfield of quantum-information thermodynamics~\cite{Lostaglio_16_Resource, NYH_18_Beyond, Guryanova_16_Thermodynamics, NYH_16_Microcanonical, Lostaglio_17_Thermodynamic, Sparaciari_20_First, Khanian_20_From, Khanian_20_Resource, Gour_18_Quantum, Manzano_22_Non, Popescu_18_Quantum, Popescu_20_Reference, NYH_16_Microcanonical, Ito_18_Optimal, Bera_19_Thermo, Mur_Petit_18_Revealing, Manzano_18_Squeezed,  NYH_20_Noncommuting, Manzano_20_Hybrid, Fukai_20_Noncommutative, Mur_Petit_20_Fluctuations, Scandi_19_Thermodynamic, Manzano_18_Squeezed, Boes_18_Statistical, Ito_18_Optimal, Mitsuhashi_22_Characterizing, Croucher_18_Information, Vaccaro_11_Information, Wright_18_Quantum, Zhang_20_Stationary, Medenjak_20_Isolated, Croucher_21_Memory, Marvian_21_Qudit, NYH_22_How, Kranzl_22_Experimental, Marvian_22_Rotationally, Ducuara_22_Quantum, Majidy_23_Non}.

\vspace{0.5em}
\emph{\textbf{Review of conventional ETH.}}---Let the Hamiltonian $H$, energy eigenstates $\ket{\alpha}$, eigenenergies $E_\alpha$, operator $\Op$, and matrix elements 
$\Op_{\alpha \alpha'} \coloneqq \bra{\alpha} \Op \ket{\alpha'}$
be defined as in the introduction.
The operator and Hamiltonian satisfy the ETH if
\begin{align}
   \label{eq_ETH}
   \Op_{\alpha \alpha'}
   = \mathcal{O}(\mathcal{E}) \, \delta_{\alpha, \alpha'}
   + e^{- \ent(\mathcal{E}) / 2} \, f(\mathcal{E}, \omega) R_{\alpha \alpha' } \, .
\end{align}
The relevant energies average to
$\mathcal{E} \coloneqq (E_{\alpha}+E_{\alpha'})/2$,
their difference is
$\omega \coloneqq E_{\alpha}-E_{\alpha'}$,
$\mathcal{O}(\mathcal{E})$ and $f(\mathcal{E}, \omega)$ 
are real functions that vary smoothly with the energy density $\mathcal{E} / \Sites$,
$\ent(\mathcal{E})$ denotes the thermodynamic entropy
(logarithm of the density of states)
at energy $\mathcal{E}$,
$\delta_{\alpha,\alpha'}$ denotes the Kronecker delta,
and the $R_{\alpha \alpha'}$ are erratically varying $O(1)$ numbers~\cite{Foini_19_ETH-OTOC, Pappalardi_22_FreeETH, Wang_22_ETH-RMT}. 
The first, ``diagonal'' ($\alpha {=} \alpha'$) term in Eq.~\eqref{eq_ETH} contains the microcanonical expectation value $\mathcal{O}(\mathcal{E})$.
The thermodynamic entropy $\ent(\mathcal{E})$ exponentially suppresses the second, ``off-diagonal'' term.

If $\Op$ and a nondegenerate $H$ satisfy the ETH, $\Op$ thermalizes~\cite{Srednicki_99_Approach, DAlessio_16_From}:
Let $\Sites$ denote the system's size.
The system begins in a normalized state
$\ket{\psi(0)}  =  \sum_\alpha  C_\alpha  \ket{\alpha}$
with an extensive energy
$E \coloneqq \langle H \rangle = O(N)$. 
We denote expectation values by 
$\langle \cdot \rangle 
\coloneqq \langle \psi(0)| \cdot | \psi(0) \rangle$.
Let the energy variance, 
$\var(H) \coloneqq \langle H^2 \rangle - E^2$, 
be at most $O(\Sites)$.

At time $t$, the operator's expectation value is
\begin{align}
   \label{eq_Exp_Val_t}
   \big\langle \Op \big\rangle_t
   = &\sum_\alpha  | C_\alpha |^2  \Op_{\alpha \alpha}
   +  \sum_{\alpha \neq \alpha'}
       C_\alpha^*  C_{\alpha'} \,
       e^{i (E_\alpha - E_{\alpha'}) t / \hbar}  \,
       \Op_{\alpha \alpha'} \, .
\end{align}
Consider averaging this value over an infinite time:
$\overline{ \expval{\Op}_t }
\coloneqq \lim_{t \to \infty}  \frac{1}{t}
\int_0^{t}  dt'  \,  \expval{ \Op }_{t'}$.
As $H$ lacks degeneracies, phase cancellations
make the second term average to zero: 
$\overline{ \expval{\Op}_t } 
= \sum_\alpha  | C_\alpha |^2  \Op_{\alpha \alpha}$.

To the first term, we apply a strategy that will echo in our noncommuting-charge arguments.
By the ETH~\eqref{eq_ETH}, 
$\Op_{\alpha \alpha} \approx \mathcal{O} (E_\alpha)$
can be Taylor-expanded about $E_\alpha = E$.
The zeroth-order term yields 
$\overline{ \expval{\Op}_t } \approx \mathcal{O}(E)$,
by the state's normalization.
The first-order term vanishes, by the definition of $E$.
All higher-order terms yield corrections $\leq O(\Sites^{-1})$, 
by the energy-variance bound and the smoothness of $\mathcal{O}(\mathcal{E})$.
Hence the time average 
$\overline{ \expval{\Op}_t }
= \mathcal{O}(E) + O(\Sites^{-1})$
approximately equals the microcanonical average.
So does the canonical average, 
$\Tr ( \Op \rho_\can) = \mathcal{O}(E) + O(\Sites^{-1})$,
by the ETH~\eqref{eq_ETH} and related arguments~\cite{Srednicki_99_Approach,Landau_Statistical_Book,Brandao_15_Equivalence,Tasaki_18_On}.
Therefore, the time average
$\overline{ \expval{\Op}_t }$ equals the thermal average 
plus $O(\Sites^{-1})$ corrections.

%
%

%
%
%
\vspace{0.5em}
\emph{\textbf{Setup suited to noncommuting charges.}}---Consider a quantum system formed from
$\Sites \gg 1$ degrees of freedom.
The Hamiltonian, $H$, is nonintegrable.
It conserves a number $\ll \Sites$ of charges $Q_a$
that do not all commute:
$[H, Q_a] = 0$, but $[Q_a, Q_{a'}] \neq 0$ for some $a' \neq a$.
The charges generate a non-Abelian symmetry group.

We illustrate with an $\Sites$-qubit system that 
has an SU(2) symmetry---whose 
total spin components $S_{a = x,y,z}$ are conserved.
(Those components decompose as
$S_a = \sum_{j=1}^\Sites  s_{j,a} \, ,$
if the $s_{j,a}$ denote qubit $j$'s spin operators.)
$H$, $\vec{S}^2$, and $S_z$ share an eigenbasis
$\{ \ket{\alpha,  m}  \}$. If $\hbar = 1$,
\begin{align}
   \label{eq_Eigenthings}
   & H  \ket{\alpha,  m}
   =  E_\alpha  \ket{\alpha,  m},  \\
   & \vec{S}^2  \ket{\alpha,  m}
   =  s_\alpha (s_\alpha + 1)  \ket{\alpha,  m}, 
   \; \; \text{and} \\ & 
   S_z  \ket{\alpha,  m}
   =  m  \ket{\alpha,  m} ,
   \; \; \text{wherein} \\ &
   \label{eq_Eigen_Bounds}
   m = - s_\alpha, - s_\alpha + 1, \ldots, s_\alpha \, .
\end{align}
Ladder operators
$S_\pm = S_x \pm i S_y$ raise and lower $S_z$

The normalized initial state decomposes as
\begin{align}
   \label{eq_Init_State}
   \ket{ \psi(0) }  =  \sum_{\alpha, m}  
   C_{\alpha, m}  \ket{\alpha, m} \, ,
   \quad \text{wherein} \quad 
   C_{\alpha, m} \in \mathbb{C} \, .
\end{align}
Operators $\Op$ have time-$t$ expectation values
$\langle \Op \rangle_t
\coloneqq \langle \psi(t) | \Op | \psi(t) \rangle$.
We drop the subscript from time constants:
\begin{align}
   \label{eq_E_def}
   \langle H \rangle &\eqqcolon E \, ,  \quad  \text{and} \\
   \label{eq_M_def}
   \langle S_z \rangle &\eqqcolon M \, .
\end{align}
Aligning the $z$-axis with $\langle \vec{S} \rangle$, we set
$M \geq 0$ and
$\langle S_x \rangle,  \langle S_y \rangle  =  0$,
without sacrificing generality.
The state has an extensive energy, $E = O(\Sites)$,
and is far from maximally spin-polarized: 
$\Sites - M = O(\Sites)$.
(ETH-type statements tend to hold when the thermodynamic entropy is extensive~\cite{DAlessio_16_From}. $\ent$ tends to be
nonextensive when additive charges [e.g., $E$ and $S_{x,y,z}$] lie near their extremes, which we therefore exclude.)

$\ket{\psi(0)}$ belongs to an \emph{approximate microcanonical subspace}, which generalizes a microcanonical subspace 
for noncommuting charges~\cite{NYH_16_Microcanonical, NYH_20_Noncommuting, Kranzl_22_Experimental}:
Measuring any charge $Q_a$ likely yields
an outcome near $\expval{ Q_a }$;
the charges' variances are bounded as
\begin{align}
   \label{eq_Var_H}
   & \var(H)  \leq  O(\Sites), \\
   \label{eq_Var_z}
   & \var(S_{z})  \leq  O(\Sites),  \quad  \text{and} \\
   \label{eq_Var_xy}
   & \var( S_{x,y} )  \leq  O(\Sites) \, .
\end{align}
Conditions~\eqref{eq_Var_H}--\eqref{eq_Var_xy} govern
typical many-body states prepared today,
including all short-range-correlated states~\footnote{
%
Let $d$ denote the spatial dimensionality.
Equations~\eqref{eq_Var_z} and \eqref{eq_Var_xy} are satisfied if spin--spin correlations 
\unexpanded{$\langle s_{j,a} s_{j'\!,a} \rangle - \langle s_{j,a} \rangle \langle s_{j'\!,a} \rangle$}
decay more quickly than $|j-j'|^{-d}$ as 
the spatial separation $|j-j'| \to \infty$.
If this latter condition governs energy-density correlations, Eq.~\eqref{eq_Var_H} holds.
}~\cite{NYH_20_Noncommuting, Kranzl_22_Experimental}.


Having introduced the initial state, we profile 
operators expected to obey the non-Abelian ETH.
Without sacrificing generality,
we focus on symmetry-adapted operators:
\emph{Spherical tensor operators} consist of components
$T^\KParen_q$ that transform irreducibly under global SU(2) rotations~\cite{Shankar_Quantum_Book}.
For example, consider an atom absorbing a photon 
(of spin $k = 1$), 
which imparts $q {=} 1$ quantum of $z$-type angular momentum.
$T^{(k {=} 1)}_{q {=} 1}$ represents the photon's effect.
Generally, the index $q = -k, -k+1, \ldots, k$.
Examples include single-spin operators:
$s_{j,z}$ is a $T^\1_0$, and 
the ladder operators
$s_{j,\pm} = s_{j,x} \pm i s_{j,y}$ are proportional to
$T^\1_{\pm 1}$ operators. 
Every operator equals a linear combination of $T^\KParen_q$ operators~\cite{Shankar_Quantum_Book}.

We focus on few-body operators, 
commonly expected to satisfy ETH-type postulates~\cite{DAlessio_16_From, Garrison_18_Does}.
More precisely, we consider 
\emph{$K$-local} operators $\Op$,
which have operator norms $\leq O(K)$. 
Examples include products of $K$ single-spin operators, e.g.,
$s_{1,x} s_{2, y} \cdots s_{K, z} + \mathrm{h.c.}$
Every $K$-local operator equals 
a linear combination of spherical-tensor components
$T^{(\leq K)}_q$.
We focus on $K = O(1)$
and hence on operators $T^\KParen_q$ with $k,q = O(1)$.

Consider representing a $T^\KParen_q$ operator as 
a matrix relative to the energy eigenbasis.
The matrix elements obey the Wigner--Eckart theorem~\cite{Shankar_Quantum_Book},
\begin{align}
   \label{eq_WigEck}
   \bra{\alpha, m}  T^\KParen_q  \ket{\alpha', m'}
   & = \braket{ s_\alpha, m }{ s_{\alpha'}, m'; k, q }
   \, \bra{\alpha} | T^\KParen | \ket{\alpha'} .
\end{align}
The first factor,
$\braket{ s_\alpha, m }{ s_{\alpha'}, m'; k, q }$,
is a \emph{Clebsch--Gordan coefficient}, 
which encodes the rules of quantum angular-momentum addition:
The coefficient is nonzero only if $m = m' + q$ and 
$s_\alpha = | s_{\alpha'} - k |, | s_{\alpha'} - k | + 1, \ldots, s_{\alpha'} + k$ 
(only if, in the photon example, 
the atomic transition obeys selection rules). 
Whereas the Clebsch--Gordan coefficient is kinematic,
the second factor in~\eqref{eq_WigEck} is dynamical.
This \emph{reduced matrix element} 
$\bra{\alpha} | T^\KParen | \ket{\alpha'}$
depends on the operator $T^\KParen_q$ and on $H$ but 
not on the quantum numbers $m$, $m'$, and $q$
(e.g., not on how many quanta of $z$-type angular momentum the photon gives the atom).

\vspace{0.5em}
\emph{\textbf{Non-Abelian ETH.}}---We now posit that the reduced matrix element
can obey the \emph{non-Abelian ETH}.
Define the average energy
$\mathcal{E} \coloneqq \frac{1}{2} (E_\alpha + E_{\alpha'})$ and
energy difference $\omega \coloneqq E_\alpha - E_{\alpha'}$.
Analogously, define the average spin quantum number
$\mathcal{S} \coloneqq \frac{1}{2} ( s_\alpha + s_{\alpha'} )$
and the difference $\nu \coloneqq s_\alpha - s_{\alpha'}$.
Denote by $\ent ( \mathcal{E}, \mathcal{S} )$ the thermodynamic entropy at 
energy $\mathcal{E}$ and spin quantum number $\mathcal{S}$.
The operator $T^\KParen_q$ and Hamiltonian $H$ obey 
the non-Abelian ETH if
\begin{align}
   \label{eq_nonAbelian_ETH}
   \bra{\alpha} | T^\KParen | \ket{\alpha'}
   & = \mathcal{T}^\KParen ( \mathcal{E}, \mathcal{S} ) \,
   \delta_{\alpha, \alpha'} 
   \\ \nonumber & \quad
   + e^{ - \ent ( \mathcal{E}, \mathcal{S} ) / 2 } \,
   f^\KParen_\nu (\mathcal{E}, \mathcal{S}, \omega)  
   R_{\alpha \alpha'} \, .
\end{align}
The real functions $\mathcal{T}^\KParen$ and $f^\KParen_\nu$ 
depend smoothly on the densities 
$\mathcal{E}/\Sites$ and $\mathcal{S}/\Sites$.
The $R_{\alpha \alpha'}$ are erraticaly varying $O(1)$ numbers, as in the conventional ETH.

Unlike $\mathcal{E}$, $\mathcal{S}$ is nonextensive, so
the $\mathcal{S}$ dependencies in~\eqref{eq_nonAbelian_ETH} may be unexpected.
Yet the Wigner--Eckart theorem~\eqref{eq_WigEck}
prevents $\bra{\alpha} | T^\KParen | \ket{\alpha'}$
from depending on $m$ or $m'$.
Hence only $\mathcal{S}$ can encode the non-Abelian-charge conservation here.

\vspace{0.5em}
\emph{\textbf{Thermal prediction.}}---Nonintegrable systems thermalize to the canonical state 
$\rho_\can  \propto e^{-\beta H}$ if just energy is conserved; 
to the grand canonical state 
$\rho_\GC  \propto  e^{- \beta (H - \mu \mathcal{N})}$ 
if the energy and particle number $\mathcal{N}$ are conserved; etc.
Which thermal state emerges depends on the charges~\cite{Landau_Statistical_Book,Jaynes_57_Information_II}.
If they fail to commute, derivations of the thermal state's form break down~\cite{NYH_18_Beyond, NYH_16_Microcanonical}.
Certain derivations were generalized in quantum-information thermodynamics to accommodate noncommuting charges~\cite{Jaynes_57_Information_II, Balian_87_Equiprobability, NYH_16_Microcanonical, Guryanova_16_Thermodynamics, Lostaglio_17_Thermodynamic}, leading to the
\emph{non-Abelian thermal state} (NATS),
\begin{align}
   \label{eq_NATS}
   \rho_\NATS 
   \coloneqq e^{ - \beta ( H - \sum_a \mu_a Q_a ) } / Z .
\end{align}
$\beta$ and the effective chemical potentials $\mu_a$ are defined by the charge expectation values,
$\Tr ( H  \rho_\NATS ) = E$ and
$\Tr ( Q_a  \,  \rho_\NATS ) 
= \langle Q_a \rangle$~\cite{NYH_20_Noncommuting}~\footnote{
%
In a non-Abelian twist on chemical potential,
the $\mu_a$ transform as an adjoint representation of SU(2).
If rotating bodies replace the spins,
the $\mu_a$ reduce to angular velocities normalized by $\beta$.
}.
The partition function is 
$Z \coloneqq \Tr ( e^{ - \beta ( H - \sum_a \mu_a Q_a ) } )$.
The NATS shares its form with the generalized Gibbs ensemble~\cite{Rigol_09_Breakdown, Rigol_07_Relaxation, Langen_15_Experimental, Vidmar_16_Generalized, Essler_16_Quench}, 
often defined for integrable Hamiltonians and usually used with commuting charges (see~\cite{Fagotti_14_On} for an exception).
Since our charges fail to commute and our $H$ is nonintegrable, we write ``NATS'' for clarity.
Signatures of $\rho_\NATS$ have emerged dynamically in numerical simulations~\cite{NYH_20_Noncommuting} and a trapped-ion experiment~\cite{Kranzl_22_Experimental},
yet full thermalization to $\rho_\NATS$ has not been observed in closed quantum systems.
Furthermore, noncommuting charges were conjectured to alter thermalization~\cite{NYH_20_Noncommuting}.

Our $z$-axis choice simplifies $\rho_\NATS$ to
$e^{ - \beta (H - \mu S_z) } / Z$
(Suppl.~Note~\ref{app_NATS_Simpl}).
Although $\rho_\NATS$ now shares its mathematical form with $\rho_\GC$,
the physics differs significantly.
Here, energy and three noncommuting charges are 
conserved globally and transported locally;
during grand canonical thermalization, 
energy and particles---two commuting charges---are.
In the grand canonical case, the global system begins in a microcanonical subspace.
Here, no nontrivial microcanonical subspace
(associated with $s_\alpha \neq 0$) exists,
and the global system begins in 
an approximate microcanonical subspace.
These differences in setup, we show,
permit differences in thermalization.

$T^\KParen_q$ has a thermal expectation value 
$\langle T^\KParen_q \rangle_\th 
\coloneqq \Tr ( T^\KParen_q \rho_\NATS )$,
whose trace we calculate using the $\ket{\alpha, m}$ basis.
We apply the Wigner--Eckart theorem~\eqref{eq_WigEck},
then the non-Abelian ETH~\eqref{eq_nonAbelian_ETH}.
The Clebsch--Gordan 
coefficient vanishes if $q \neq 0$, so
\begin{align}
   \label{eq_Therm_ExpVal1}
   \big\langle T^\KParen_q \big\rangle_\th
   & = \frac{ \delta_{q,0} }{Z}  
   \sum_{\alpha, m}
   e^{-\beta (E_\alpha - \mu m) }  \,
   \braket{s_\alpha, m}{s_\alpha, m; k, 0} 
   \nonumber \\ 
   & \qquad \qquad \quad \; 
   \times
   \mathcal{T}^\KParen ( E_\alpha, s_\alpha ) \, .
\end{align}
(We omit corrections exponentially small in $\Sites$.)

\vspace{0.5em}
\emph{\textbf{Time-averaged expectation value.}}---After $\ket{\psi(0)}$ [Eq.~\eqref{eq_Init_State}] 
evolves for a time $t$,
the operator $T^\KParen_q$ has an expectation value
\begin{align}
   \label{eq_Time_t_ExpVal}
   \big\langle T^\KParen_q \big\rangle_t
   & = \sum_{ \alpha, \alpha', m, m' }
   C_{\alpha, m}^*  C_{\alpha', m'}  \,
   e^{i (E_\alpha - E_{\alpha'}) t } 
   \\ & \qquad \qquad \qquad \times \nonumber
   \bra{\alpha, m}  T^\KParen_q  \ket{\alpha', m'} .
\end{align}
We apply the Wigner--Eckart theorem~\eqref{eq_WigEck},
invoke the non-Abelian ETH~\eqref{eq_nonAbelian_ETH},
   %
and average $\langle T^\KParen_q \rangle_{t'}$ over an infinite time
($\lim_{t \to \infty}  \frac{1}{t}  \int_0^t dt'$). 
For all $\alpha' \neq \alpha$,
the exponential in~\eqref{eq_Time_t_ExpVal} dephases,
so the ``off-diagonal'' terms vanish:
\begin{align}
   \label{eq_Time_Avg_NATS}
   \overline{ \big\langle T^\KParen_q \big\rangle_t }
   & = \sum_{\alpha, m} 
   C_{\alpha, m + q}^*  C_{\alpha, m}
   \braket{s_\alpha, m + q}{s_\alpha, m; k, q} 
   \nonumber \\ 
   & \qquad \qquad 
   \times
   \mathcal{T}^\KParen (E_\alpha, s_\alpha) \, .
\end{align}

\vspace{0.5em}
\emph{\textbf{Comparison.}}---We prove two results:
(i) If $M = O(\Sites)$, the time average~\eqref{eq_Time_Avg_NATS} equals the thermal average~\eqref{eq_Therm_ExpVal1},
plus $O(\Sites^{-1})$ corrections, as
in the absence of noncommuting charges~\cite{Srednicki_96_Thermal, Srednicki_99_Approach}.
(ii) If $M = 0$, the time average may deviate from 
the thermal average by anomalously large,
$O(\Sites^{-1/2})$ corrections.
These corrections appear sourced by different physics:
quantum uncertainty in noncommuting charges, rather than
thermodynamic ensembles' distinguishability 
at finite $\Sites$~\cite{Srednicki_96_Thermal, Srednicki_99_Approach}.
Result (ii) holds under a physically reasonable assumption 
described and motivated below Eq.~\eqref{eq_Time_Avg_Anom}.
Anomalous thermalization may occur also at intermediate scalings $M = O(N^{\gamma})$, for exponents $0 < \gamma < 1$, 
but this regime lies outside this paper's scope.

Consider an extensive $M = O(\Sites)$ and 
$s_{j,z}$-like operators $T^\KParen_{q=0}$.
We sketch the argument for thermalization here; details appear in 
Suppl.~Note~\ref{app_Prove_Therm_q0}.
The thermal average~\eqref{eq_Therm_ExpVal1} 
and time average~\eqref{eq_Time_Avg_NATS}
share a crucial property: In each, 
$\mathcal{T}^\KParen (E_\alpha, s_\alpha)$
is averaged over a sharply peaked probability distribution.
The peaking follows primarily from the variance conditions~\eqref{eq_Var_H}--\eqref{eq_Var_xy}.
Near each peak, the smooth function 
$\mathcal{T}^\KParen (E_\alpha, s_\alpha)$ 
can be Taylor-expanded, then averaged term by term.
The leading term evaluates to $\mathcal{T}^\KParen (E, M)$ 
in both averages,~\eqref{eq_Therm_ExpVal1} and~\eqref{eq_Time_Avg_NATS}.
All higher-order terms evaluate to 
$\leq O(\Sites^{-1})$.
Therefore, the averages equal each other
to within the usual correction:
\begin{align}
   \label{eq_p_Avg_q0_b}
   \overline{ \big\langle T^\KParen_0 \big\rangle_t }
   - \big\langle T^\KParen_0 \big\rangle_\th
   = O(\Sites^{-1}) .
\end{align}

Now, consider ladder-operator-like operators
$T^\KParen_{q \neq 0}$.
The thermal average~\eqref{eq_Therm_ExpVal1} vanishes,
due to the Kronecker delta.
The time average~\eqref{eq_Time_Avg_NATS} is
$\leq O(\Sites^{-1})$,
as shown in 
Suppl.~Notes~\ref{app_Time_Avg_q_neq_0} and~\ref{app_Approx_CGs}.
Hence the time average equals the vanishing thermal average
to within the ordinary $O(\Sites^{-1})$ correction. 

The correction can be anomalously large when $M = 0$.
When $M = 0$, the thermal state is rotationally invariant.
Only similarly invariant $T^\0_0$ operators can have 
nonzero thermal averages~\footnote{
%
One can check this claim using Eq.~\eqref{eq_Therm_ExpVal1}.
Since $M = 0$, $\mu = 0$, so the $\sum_m$ vanishes if $k > 0$.
}.
Contrariwise, some states $\ket{\psi(0)}$ have $M=0$ but are rotationally \emph{noninvariant}. 
Intuitively, these states have
vanishing magnetic dipole moments but nonzero magnetic quadrupole moments (or higher-order moments).
Such states can endow operators $T^{(k>0)}_q$ with 
time averages of $O(\Sites^{-1/2})$,
in contrast with their vanishing thermal averages.
Here is an example.

Consider an arbitrary Hamiltonian eigenspace labeled by 
$\alpha = A$, associated with an extensive energy 
$E_A = O(\Sites)$ and a spin quantum number
$s_A = O(\Sites^{1/2})$
(chosen for reasons shown below).
The following state has $M=0$ but is rotationally noninvariant:
\begin{align}
   \label{eq_Init_State_Anom}
   \ket{\psi(0)}
   \coloneqq \sqrt{\frac{1}{3}} \: \ket{A, m{=}s_A} 
   + \sqrt{\frac{2}{3}} \: 
   \bigg\lvert A, m{=}-\frac{s_A}{2} \bigg\rangle .
\end{align}
$\ket{\psi(0)}$ has the properties stipulated in the setup,
one can check directly.
Consider the local magnetic quadrupole moment
$3 s_{i,z} s_{j,z} - \vec{s}_i \cdot \vec{s}_j$.
The $i$ and $j$ label neighboring sites.
This $T^\2_0$ operator's time average~\eqref{eq_Time_Avg_NATS} reduces to
\begin{align}
   \label{eq_Time_Avg_Anom}
   \overline{ \big\langle T^\2_0 \big\rangle_t }
   = O(1) \times
   \mathcal{T}^\2 (E_A, s_A) .
\end{align}
Clebsch--Gordan coefficients determine the $O(1)$ factor.

The $\mathcal{T}^\2 (E_A, s_A)$ scales
linearly with the spin density---as $s_A / \Sites$---for some systems, we assume.  
Bound states motivate this assumption, as 
outlined here and detailed in 
Suppl.~Note~\ref{app_Form_of_Tk}.
$\mathcal{T}^\2 (E_A, s_A)$
approximately equals an eigenstate expectation value,
by the Wigner--Eckart theorem and the non-Abelian ETH:
\begin{align}
   \mathcal{T}^\2 (E_A, s_A) 
   \approx \bra{A, s_A} 
   3 s_{i,z} s_{j,z} - \vec{s}_i \cdot \vec{s}_j 
   \ket{A, s_A} .
\end{align}
The right-hand side is essentially 
the joint probability $P(i, j)$ of finding 
spin quanta at sites $i$ and $j$.
Semiclassically, $P(i,j) = P(i | j) \times P(j)$, if $P(j)$ denotes the probability of finding a quantum at $j$ and $P(i|j)$ denotes the conditional probability of then finding a quantum at $i$.
$P(i|j)$ can be $O(1)$ if the spin quanta form bound clusters:
Just as attractive interactions can bind particles together, so may suitable (e.g., ferromagnetic) couplings bind spin quanta.
In the high-energy eigenstate $\ket{A,s_A}$, 
clusters will be spread uniformly, 
with a density $\sim s_A / \Sites \sim P(j)$.
Combining these steps yields 
$\mathcal{T}^\2 (E_A, s_A) 
\sim P(i | j) \times P(j) 
= O(1) \times O(s_A / \Sites) = O(\Sites^{-1/2})$, 
by our choice $s_A = O( N^{1/2} )$.
Substituting into Eq.~\eqref{eq_Time_Avg_Anom}
yields the time average. 
It deviates from the vanishing thermal average by
\begin{align}
   \label{eq_Correction_Anom}
   \overline{ \big\langle T^\2_0 \big\rangle_t }
   - \big\langle T^\2_0 \big\rangle_\th
   = O(\Sites^{-1/2}) > O(\Sites^{-1}) .
\end{align}
See 
Suppl.~Note~\ref{app_Anomaly2} 
for details and 
Suppl.~Note~\ref{app_T00_Avgs} 
for another anomalous-thermalization example.

Anomalous $O( \Sites^{-1/2} )$ scaling
characterizes also a kinematic bound in Ref.~\cite{NYH_16_Microcanonical}.
That work generalized
a conventional derivation of the thermal state's form
to accommodate noncommuting charges:
The global system, formed from 
$\Sites$ identical subsystems, was assumed to be in 
a generalized microcanonical state.
The average subsystem's reduced state was found to lie a distance
$\leq (\const) \Sites^{-1/2}  +  (\const)$
from $\rho_\NATS$.
It is possible that our results, 
based on dynamics and the ETH,
reflect the Hamiltonian-independent results in~\cite{NYH_16_Microcanonical}.

\vspace{0.5em}
\emph{\textbf{Outlook.}}---We have extended the eigenstate thermalization hypothesis, a cornerstone of many-body physics, to the more fully quantum scenario in which conserved charges fail to commute with each other. Noncommutation can prevent the charges from sharing an eigenspace (a sector) and invalidates the usual assumption 
of the Hamiltonian's nondegeneracy.
We overcame these challenges by proposing a non-Abelian ETH and focusing on an approximate microcanonical subspace. Applying these tools to SU(2), we compared the long-time average of an operator's expectation value with the thermal expectation value.
The averages agree in many cases, e.g., whenever $M = O(\Sites)$.
Yet the averages can disagree by 
anomalously large $O(\Sites^{-1/2})$ corrections
under a physically reasonable assumption.

This work establishes several research opportunities.
First, our analytical results call for 
testing with numerics and quantum simulators.
Trapped ions have been shown, and ultracold atoms and superconducting qudits have been argued, to be able to test noncommuting-charge thermodynamics~\cite{NYH_20_Noncommuting,NYH_22_How,Kranzl_22_Experimental}.
Promising models include nonintegrable Heisenberg Hamiltonians~\cite{NYH_20_Noncommuting,NYH_22_How,Kranzl_22_Experimental} and many-electron atoms.
One would verify the non-Abelian ETH~\eqref{eq_nonAbelian_ETH}; 
identify operators $T^\KParen_q$ whose smooth functions $\mathcal{T}^\KParen$ satisfy our assumptions, 
enabling anomalous thermalization; and observe deviations~\eqref{eq_Correction_Anom} from thermal predictions.

Second, those deviations may signal the retention, by local subsystems, of information about their initial conditions. 
Such retention might be leveraged. 
Noncommuting charges could enhance quantum memories, as many-body localization has been proposed to~\cite{Nandishore_15_Many}.
Localization resembles prethermalization~\cite{Mori_18_Thermalization}, scars~\cite{Chandran_22_Scars-Review}, and Hilbert-space fragmentation~\cite{Moudgalya_22_Quantum} in disrupting closed quantum many-body systems' thermalization. Noncommuting charges may belong on the list, our results indicate. Confirmation would hold fundamental interest, as disrupting thermalization effectively hinders time's arrow. 

Third, our arguments merit generalization from SU(2).
Fourth, the smooth function 
$f^\KParen_\nu( \mathcal{E}, \mathcal{S}, \omega)$ [Eq.~\eqref{eq_nonAbelian_ETH}] should reveal how
non-Abelian symmetries influence thermalization \emph{dynamics} and so merits investigation.
This work extends the ETH to 
the more fully quantum regime of noncommuting charges,
linking many-body physics to quantum-information thermodynamics~\cite{Lostaglio_16_Resource, NYH_18_Beyond, Guryanova_16_Thermodynamics, NYH_16_Microcanonical, Lostaglio_17_Thermodynamic, Sparaciari_20_First, Khanian_20_From, Khanian_20_Resource, Gour_18_Quantum, Manzano_22_Non, Popescu_18_Quantum, Popescu_20_Reference, NYH_16_Microcanonical, Ito_18_Optimal, Bera_19_Thermo, Mur_Petit_18_Revealing, Manzano_18_Squeezed,  NYH_20_Noncommuting, Manzano_20_Hybrid, Fukai_20_Noncommutative, Mur_Petit_20_Fluctuations, Scandi_19_Thermodynamic, Manzano_18_Squeezed, Boes_18_Statistical, Ito_18_Optimal, Mitsuhashi_22_Characterizing, Croucher_18_Information, Vaccaro_11_Information, Wright_18_Quantum, Zhang_20_Stationary, Medenjak_20_Isolated, Croucher_21_Memory, Marvian_21_Qudit, NYH_22_How, Kranzl_22_Experimental, Marvian_22_Rotationally, Ducuara_22_Quantum, Majidy_23_Non}.

%
%
\begin{acknowledgments}
C.M.~thanks Tibor Rakovszky and Curt von Keyserlingk for helpful conversations; and N.Y.H.~thanks Wen Wei Ho, Noah Lupu-Gladstein, Mikhail Lukin, and Bihui Zhu. 
C.M.~and N.Y.H.~acknowledge the hospitality of the Kavli Institute for Theoretical Physics, where this work started.
N.Y.H.~acknowledges also the Perimeter Institute's hospitality.
This research was supported in part by the Gordon and Betty Moore Foundation's EPiQS Initiative through GBMF8686 (C.M.),
by the National Science Foundation under QLCI grant OMA-2120757 (N.Y.H.), 
by an NSF Graduate Research Fellowship under Grant No.~2139319 (F.I.), 
and by the National Science Foundation under Grant No.~NSF PHY-1748958 (C.M., M.S., and N.Y.H.). 
Research at Perimeter Institute is 
supported in part by the Government of Canada through the Department of Innovation, Science and Economic Development Canada and by the Province of Ontario through the Ministry of Colleges and Universities.
\end{acknowledgments}

%

%

%

\clearpage

\onecolumngrid

\begin{center}

{\large \bf Supplemental Material for\\[0.2em]
``Non-Abelian eigenstate thermalization hypothesis''}\\[0.8em]

Chaitanya~Murthy,${}^1$ 
Arman~Babakhani,${}^{2,3}$
Fernando~Iniguez,${}^4$
Mark~Srednicki,${}^4$
and Nicole~Yunger~Halpern${}^{5,6}$\\[0.5em]
{\small \it
${}^1$Department of Physics, Stanford University, Stanford, CA 94305, USA\\[0.05em]
${}^2$Department of Physics, University of Southern California, Los Angeles, CA, 90089, USA\\[0.05em]
${}^3$Information Sciences Institute, Marina Del Rey, CA, 90292, USA\\[0.05em]
${}^4$Department of Physics, University of California, Santa Barbara, CA 93106, USA\\[0.05em]
${}^5$Joint Center for Quantum Information and Computer Science,\\ NIST and University of Maryland, College Park, MD 20742, USA\\[0.05em]
${}^6$Institute for Physical Science and Technology, University of Maryland, College Park, MD 20742, USA}

\end{center}

\vspace{0em}

\renewcommand{\thesection}{\arabic{section}}
\renewcommand{\thesubsection}{\Alph{subsection}}

\setcounter{equation}{0}
\renewcommand{\theequation}{S\arabic{equation}}

\setcounter{footnote}{0}
\renewcommand{\thefootnote}{S\arabic{footnote}}

%

%
%
%
\section{Simplification of the non-Abelian thermal state}
\label{app_NATS_Simpl}

The non-Abelian thermal state,
$\rho_\NATS \coloneqq e^{ - \beta ( H - \sum_a \mu_a Q_a ) } / Z \, ,$
appears as Eq.~\eqref{eq_NATS} in the main text.
In the setup section, we chose for the $\hat{z}$-axis to align with the magnetization, such that $\expval{ S_{x,y} } = 0$. 
This choice, we claimed, simplifies the mathematical form of $\rho_\NATS$ to 
$e^{ - \beta (H - \mu S_z) } / Z \, .$
We prove that claim here.

Conserving $S_{x,y,z} \, ,$ $H$ shares an eigenbasis with every linear combination thereof,
including $\vec{\mu} \cdot \vec{S} 
= \mu (\hat{\mu} \cdot \vec{S}) \, ,$
wherein $\mu \coloneqq || \vec{\mu} || \, .$
Denote such a shared eigenbasis by
$\{ \ket{\alpha, \tilde{m} } \}$:
$H \ket{\alpha, \tilde{m} } 
= E_\alpha \ket{\alpha, \tilde{m} } \, ,$
$\vec{S}^2 \ket{\alpha, \tilde{m} }
= s_\alpha (s_\alpha + 1) \ket{\alpha, \tilde{m} } \, ,$ and
$\vec{\mu} \cdot \vec{S}
\ket{\alpha, \tilde{m} }
= \mu \tilde{m} \ket{\alpha, \tilde{m} } \, .$
Since $\hat{\mu} \cdot \vec{S}$
is a component of $\vec{S} \, ,$
by the ordinary algebra of quantum angular momentum,
$\tilde{m} = -s_\alpha, -s_\alpha + 1, 
\ldots, s_\alpha \, .$

Using the shared eigenbasis, we calculate the thermal expectation value
\begin{align}
   \expval{ \vec{S} }_\th
   & = \frac{1}{Z} \, \Tr \left( \vec{S} \,
   e^{ - \beta (H - \vec{\mu} \cdot \vec{S}) } \right)
   = \frac{1}{Z} \sum_{\alpha, \tilde{m} }
   \bra{\alpha, \tilde{m} } \vec{S} \, 
   e^{-\beta (H - \vec{\mu} \cdot \vec{S}) } \,
   \ket{\alpha, \tilde{m} }
   \label{eq_Simpl_NATS_Help1}
   = \frac{1}{Z} \sum_{\alpha, \tilde{m} }
   \bra{\alpha, \tilde{m} } \vec{S} 
   \ket{\alpha, \tilde{m} } \,
   e^{-\beta (E_\alpha - \mu \tilde{m} ) } \, .
\end{align}
The final equality follows from the eigenvalue equations.
To calculate $\bra{\alpha, \tilde{m} } \vec{S} 
\ket{\alpha, \tilde{m} } \, ,$
imagine that $m$'s replaced the $\tilde{m}$'s.
The inner product would be
$\bra{\alpha, m} \vec{S} \ket{\alpha, m}
= \bra{\alpha, m} S_z \hat{z} \ket{\alpha, m}
= m \hat{z} \, .$
Analogously,
$\bra{\alpha, \tilde{m} } \vec{S} 
\ket{\alpha, \tilde{m} }
= \bra{\alpha, \tilde{m} }
\hat{\mu} \cdot \vec{S}
\ket{\alpha, \tilde{m} }
= \tilde{m} \hat{\mu} \, .$
Hence~\eqref{eq_Simpl_NATS_Help1} reduces to
\begin{align}
   \label{eq_Simpl_NATS_Help2}
   \expval{ \vec{S} }_\th
   & = \frac{ \hat{\mu} }{Z} \sum_{\alpha, \tilde{m} }
   \tilde{m} \,
   e^{-\beta (E_\alpha - \mu \tilde{m}) } 
   = \frac{ \hat{\mu} }{Z} \sum_{\alpha, m} m \,
   e^{-\beta (E_\alpha - \mu m) }
   = \frac{ \hat{\mu} }{Z} \,
   \Tr\!\left( S_z \,
   e^{-\beta (H - \mu S_z) } \right) \, .
\end{align}
The second equality follows from the $\tilde{m}$ values' being the same as the $m$ values.
Hence $\langle \vec{S} \rangle_\th$ points in
the direction $\hat{\mu} \, .$ Therefore,
if $\langle S_{x,y} \rangle = 0$, then
$\mu_{x,y} = 0 \, .$

\section{Time-averaged expectation values thermalize to within $O(N^{-1})$ corrections if \texorpdfstring{$M = O(\Sites)$}{M = O(\Sites)} and \texorpdfstring{$q = 0$}{q = 0}}
\label{app_Prove_Therm_q0}

This Suppl.~Note details the thermalization proof 
sketched in the main text.
Let the magnetization be extensive, $M = O(\Sites)$, 
and consider $s_{j,z}$-like operators $T^\KParen_{q=0}$.
The thermal average~\eqref{eq_Therm_ExpVal1} 
and time average~\eqref{eq_Time_Avg_NATS} both assume the form
\begin{align}
   \label{eq_p_Avg_q0_a}
   \sum_{\alpha, m}
   p_{\alpha, m}
   \braket{s_\alpha, m }{s_\alpha, m; k, 0} \,
   \mathcal{T}^\KParen (E_\alpha, s_\alpha) 
   \eqqcolon \expval{ T^\KParen_0 }_{p} .
\end{align}
The distribution $\{ p_{\alpha, m} \}$ equals
$\{ e^{-\beta (E_\alpha - \mu m) } / Z \}$
in the thermal average and,
in the time average, equals the diagonal ensemble
$\{ | C_{\alpha, m} |^2 \}$.
The probabilities $p_{\alpha, m}$ 
are unit-normalized and have bounded moments:
For all nonvanishing triples 
$(A, B, C) \in 
   (\mathbb{Z}_{\geq 0})^3 \setminus (0, 0, 0)$
of non-negative integers,
\begin{align}
   \label{eq_Moment_Condn} &
   \expval{ (E_\alpha - E)^A \, 
   (m - M)^B \,
   (s_\alpha - M)^C }_p
   \leq O \left( N^{A + B + C - 1} \right) .
\end{align}
For a variable $X$, we have defined $\langle X \rangle_p$ as 
the average over $\{ p_{\alpha, m} \}$.
The thermal and diagonal distributions satisfy the moment condition~\eqref{eq_Moment_Condn} by the scalings
$E, M = O(\Sites)$ and 
the variance conditions~\eqref{eq_Var_H}--\eqref{eq_Var_xy}.
We prove this claim for the initial state in Suppl.~Note~\ref{app_Moment_Condn_Init}.
$\rho_\NATS$ satisfies the variance conditions~\eqref{eq_Var_H}--\eqref{eq_Var_xy} by
Laplace's method and, independently, by the thermal state's being short-range-correlated~\cite{Kliesch_18_Properties}.

Using the moment condition~\eqref{eq_Moment_Condn},
we evaluate the average~\eqref{eq_p_Avg_q0_a}.
We outline the calculation here; 
Suppl.~Note~\ref{app_q0} contains details.
Consider the summand in~\eqref{eq_p_Avg_q0_a}.
The probability $p_{\alpha, m}$ peaks about
$(E_\alpha = E, \, m = M, \, s_\alpha = M)$,
by the moment condition~\eqref{eq_Moment_Condn}.
In contrast, the Clebsch--Gordan coefficient and 
$\mathcal{T}^\KParen$ are smooth.
They can therefore be Taylor-expanded about the peak.
In the Taylor expansion, a general term is 
an $n^\th$-order derivative times an $n^\th$-order moment,
for some $n \geq 0$. 
The derivative is $\leq O(\Sites^{-n})$, by the functions' smooth dependence on $E_{\alpha}/\Sites$, $m/\Sites$, and $s_{\alpha}/\Sites$.
By the moment condition~\eqref{eq_Moment_Condn}, 
the moment is $\leq O(\Sites^{n-1})$. Hence
Eq.~\eqref{eq_p_Avg_q0_a} reduces to
\begin{align}
   \label{eq_p_Avg_q0_b_app}
   \expval{ T^\KParen_0 }_p
   = \mathcal{T}^\KParen (E, M) + O(\Sites^{-1}) \, .
\end{align}
The left-hand side equals the thermal average, for 
one instance of the probability distribution $p$,
as well as the time average, for another instance.
Hence the averages equal each other to within 
the usual $O(\Sites^{-1})$ corrections.

\subsection{The initial state satisfies the moment condition}
\label{app_Moment_Condn_Init}

The previous section casts the diagonal ensemble 
$\{ | C_{\alpha, m} |^2 \}$ as satisfying
the moment condition~\eqref{eq_Moment_Condn}.
We prove the claim here.
The proof relies on three ingredients: 
(i) the finite dimensionality of the local subsystems' Hilbert spaces;
(ii) the scalings 
\begin{align}
   \label{eq_Scalings_App}
   E = O(\Sites)
   \quad \text{and} \quad
   M = O(\Sites) \, ;
\end{align}
and (iii) the variance conditions~\eqref{eq_Var_H}--\eqref{eq_Var_xy}, 
repeated here for convenience:
\begin{align}
   \label{eq_Var_H_App}
   & \var(H) = \langle H^2 \rangle - E^2  \leq  O(\Sites), \\
   \label{eq_Var_z_App}
   & \var(S_{z}) = \langle S_z^2 \rangle - M^2 \leq  O(\Sites),  \quad  \text{and} \\
   \label{eq_Var_xy_App}
   & \var( S_{x,y} ) = \langle S_{x,y}^2 \rangle \leq  O(\Sites) \, .
\end{align}
Recall that the variance conditions are satisfied, for example, if $\ket{\psi(0)}$ is short-range-correlated.

The proof proceeds in three steps.
First, we derive analogs for $\vec{S}^2$ of
the scaling and variance conditions~\eqref{eq_Scalings_App}--\eqref{eq_Var_xy_App}.
Second, we upper-bound fairly general correlators' magnitudes.
Third, we combine steps 1--2.

\vspace{1em}
\emph{Step 1:} Unlike $S_z$, $\vec{S}^2$ is nonextensive.
Therefore, $\vec{S}^2$'s expectation value and variance
are not necessarily bounded as $S_z$'s are.
However, we can prove similar bounds.
We define the diagonal average of any variable $X_{\alpha,m}$ as
\begin{align}
    \langle X_{\alpha, m} \rangle_\diag
    \coloneqq \sum_{\alpha, m} 
    | C_{\alpha, m} |^2 X_{\alpha, m} .
\end{align}
We will prove the bounds
\begin{align}
   \label{eq_Bound_S2_App}
   \expval{s_\alpha}_\diag \leq M + O(1)
   \quad \text{and} \quad
   \expval{ (s_\alpha - M)^2 }_\diag
   \leq O(\Sites) \, .
\end{align}

First, we sum the variance conditions~\eqref{eq_Var_z_App} and~\eqref{eq_Var_xy_App}: 
\begin{align}
   \langle \vec{S}^2 \rangle - M^2 \leq O(N) .
\end{align}
We evaluate the left-hand side (LHS) on $\ket{\psi(0)}$ [Eq.~\eqref{eq_Init_State}]:
\begin{align}
   \label{eq_sm_bound_App}
   \sum_{\alpha, m} | C_{\alpha, m} |^2 s_\alpha (s_\alpha + 1) - M^2 \leq O(\Sites) .
\end{align}
The inequality is equivalent, 
by algebra and the normalization of
$\{ | C_{\alpha, m} |^2 \}$, to
\begin{align}
   \label{eq_Bound_s_Help1_App}
   M  + (2 M + 1) \sum_{\alpha, m} 
   | C_{\alpha, m} |^2 (s_{\alpha} - M)
   + \sum_{\alpha, m} | C_{\alpha, m} |^2
   (s_{\alpha} - M)^2
   \leq O(\Sites) \, .
\end{align}
Since $M = \sum_{\alpha, m} |C_{\alpha, m}|^2 \, m \, ,$
we can replace the second term's 
$(s_{\alpha} - M)$ with $(s_{\alpha} - m)$.
Recall that $m \leq s_{\alpha}$.
Every factor on the inequality's LHS is therefore nonnegative, 
so every term is,
so every term must be $\leq O(\Sites) \, .$
Since $2M+1 = O(\Sites)$, the second term implies that
$\sum_{\alpha, m} | C_{\alpha, m} |^2 (s_\alpha - M) \leq O(1) \, .$
By the definition of $\langle . \rangle_\diag$, we recover the first inequality in~\eqref{eq_Bound_S2_App}.
The second inequality in~\eqref{eq_Bound_S2_App} follows similarly from
Ineq.~\eqref{eq_Bound_s_Help1_App}'s third term.

\vspace{1em}
\emph{Step 2:} We now upper-bound fairly general correlators' magnitudes.
Let $x_1, x_2, \ldots, x_n$ denote real-valued functions of 
$\alpha$ and $m$. 
(For notational brevity, we suppress the functions' dependencies on $\alpha$ and $m$.)
Let the functions' magnitudes obey the upper bound
$| x_j | \leq X \in \mathbb{R} \; \: \forall \, j, \alpha, m$.
We analyze correlator magnitudes of the form
\begin{align}
   \label{eq_Corr_Mag_App}
   \left\lvert \expval{ x_1^{A_1} x_2^{A_2} \ldots x_n^{A_n} 
   }_\diag \right\rvert \, .
\end{align}
Without loss of generality, the powers are ordered from greatest to least: $A_1 \geq A_2 \geq \ldots \geq A_n \geq 0$.
At-least-two-point correlators interest us, so
$A \coloneqq \sum_{j=1}^n A_j \geq 2$. 
Therefore, either $A_1 \geq 2$ or $A_1 = A_2 = 1$.
In the first case, we show, the correlator magnitude~\eqref{eq_Corr_Mag_App} is upper-bounded by
$X^{A-2}$ times $\langle x_1^2 \rangle$; 
in the second case, the correlator magnitude is upper-bounded by $X^{A-2}$ times $\frac{1}{2} \langle x_1^2 + x_2^2 \rangle$.
We parcel the factors so for reasons clarified in step 3.

First, suppose that $A_1 \geq 2$.
To upper-bound~\eqref{eq_Corr_Mag_App}, 
we invoke the average's definition, then the triangle inequality:
\begin{align}
   \left\lvert \expval{ x_1^{A_1} x_2^{A_2} \ldots x_n^{A_n} 
   }_\diag \right\rvert
   \leq \sum_{\alpha, m} 
   \left\lvert C_{\alpha, m} \right\rvert^2
   |x_1|^{A_1}
   |x_2|^{A_2}
   \cdots
   |x_n|^{A_n} .
\end{align}
We separate out a factor of $|x_1|^2$.
Then, we bound the rest using the assumption $|x_j| \leq X$ and
the definition $A \coloneqq \sum_{j=1}^n A_j$:
\begin{align}
   \label{eq_Corr_Mag_b_App}
   \left\lvert \expval{ x_1^{A_1} x_2^{A_2} \ldots x_n^{A_n} 
   }_\diag \right\rvert
   & \leq \sum_{\alpha, m} 
   \left\lvert C_{\alpha, m} \right\rvert^2
   |x_1|^2 \cdot
   \underbrace{ |x_1|^{A_1-2} |x_2|^{A_2} 
   \cdots |x_n|^{A_n}
   }_{ \leq X^{A-2} }
   \leq \expval{ x_1^2 }_\diag  X^{A-2} \, .
\end{align}
The final inequality follows from the reality of $x_1$.

Now, suppose that $A_1 = A_2 = 1$.
To bound the correlator magnitude~\eqref{eq_Corr_Mag_App},
we again invoke the average's definition, 
then the triangle inequality.
This time, we separate $x_1^{A_1} x_2^{A_2} = x_1 x_2$ from the other variables:
\begin{align}
   \left\lvert \expval{ x_1^{A_1} x_2^{A_2} \ldots x_n^{A_n} 
   }_\diag \right\rvert
   \label{eq_Corr_Mag_c_App}
   & \leq \sum_{\alpha, m}
   \left\lvert C_{\alpha, m} \right\rvert^2
   | x_1 x_2 | \cdot
   \underbrace{ |x_3|^{A_3} |x_4|^{A_4} 
   \cdots |x_n|^{A_n}
   }_{ \leq X^{A-2} } \, .
\end{align}
Since $x_1$ and $x_2$ are real,
$x_1^2 + x_2^2 - 2 |x_1 x_2| = (|x_1| - |x_2|)^2 \geq 0$.
Rearranging yields $|x_1 x_2| \leq \frac{1}{2} (x_1^2 + x_2^2)$.
Combining this inequality with Ineq.~\eqref{eq_Corr_Mag_c_App},
we obtain
\begin{align}
   \left\lvert \expval{ x_1^{A_1} x_2^{A_2} \ldots x_n^{A_n} 
   }_\diag \right\rvert
   \label{eq_Corr_Mag_d_App}
   \leq \frac{1}{2} 
   \expval{x_1^2 + x_2^2}_\diag
   X^{A - 2} \, .
\end{align}

\vspace{1em}
\emph{Step 3:} We now synthesize steps 1 and 2.
Let $(x_1, x_2, x_3)$ equal 
$(E_\alpha - E, \, m - M, \, s_\alpha - M)$
or some permutation thereof.
Since local subsystems have finite-dimensional Hilbert spaces, each variable is upper-bounded by some $O(\Sites)$ number $X$.
By the variance conditions, the functions
$\langle x_j^2 \rangle_\diag$ and
$\frac{1}{2} \langle x_j^2 + x_k^2 \rangle_\diag$ are
$O(\Sites)$ for all $j, k = 1, 2, 3 \, .$
Therefore, substituting into Eq.~\eqref{eq_Corr_Mag_b_App} yields 
the moment condition~\eqref{eq_Moment_Condn},
as does substituting into Eq.~\eqref{eq_Corr_Mag_d_App}.

We can now explain why, during step 2,
we sought bounds that contained 
$\langle x_1^2 \rangle_\diag$
or $\langle x_1^2 + x_2^2 \rangle_\diag$.
These averages are only $O(\Sites)$.
If we had treated $x_1^2$ or $x_1 x_2$ like the other variables,
each would have contributed an $O(\Sites^2)$ factor to 
the corresponding bound.
We would not have recovered the all-important $-1$ in the moment condition's exponent [Eq.~\eqref{eq_Moment_Condn}].

\subsection{Detailed calculation of general average}
\label{app_q0}

The average~\eqref{eq_p_Avg_q0_a} equals 
the thermal expectation for one distribution $\{ p_{\alpha, m} \}$
and, for another distribution, equals the time-averaged expectation value.
We evaluate~\eqref{eq_p_Avg_q0_a} in detail here.
The moment condition~\eqref{eq_Moment_Condn} implies that
$\{ p_{\alpha, m} \}$ peaks near 
$(E_\alpha = E, \, m = M, \, s_\alpha = M) \, .$
Furthermore, the Clebsch--Gordan coefficient and
$\mathcal{T}^\KParen$ vary slowly.
Hence we can Taylor-expand each function about the peak. 
We do so consecutively, then combine both approximations in the sum~\eqref{eq_p_Avg_q0_a}.

\vspace{1em}
\emph{Taylor expansion of Clebsch--Gordan coefficient:}
In Suppl.~Note~\ref{app_Approx_CGs}, we approximate
the Clebsch--Gordan coefficients at $s_\alpha \gg 1$ and
$s_\alpha - m \ll s_\alpha \, .$
The latter condition does govern the dominant contributions to Eq.~\eqref{eq_p_Avg_q0_a}: 
The moment condition~\eqref{eq_Moment_Condn}, together with $M = O(N) \, ,$ implies that 
$\langle (s_\alpha - m)^n/s_\alpha^n \rangle_p \leq O(N^{-1}) \, ,$ for $n \geq 1 \, .$
To prove this inequality, we Taylor-expand 
$(s_\alpha - m)^n/s_\alpha^n$ about $m = M$ and $s_\alpha = M \, .$
A general term in the expansion is of
$O\! \left( [m - M]^B [s_\alpha - M]^C / M^{B+C} \right) \, ,$ 
wherein $B, C \in \mathbb{Z}_{\geq 0}$ and $B+C \geq n \, .$
Each such term averages to $\leq O(\Sites^{-1}) \, .$ Hence
\begin{align}
   \label{eq_sm_Moment_App}
   \left\langle \bigg[ \frac{s_\alpha - m}{s_\alpha} \bigg]^n \right\rangle_{\!p}
   \leq O\! \left(\Sites^{-1} \right) \, , 
   \quad \text{for} \quad 
   n \geq 1.
\end{align}

Having justified the use of the asymptotic expansion in Suppl.~Note~\ref{app_Approx_CGs}, we now use the expansion.
Substituting $q=0$ into Eq.~\eqref{eq_CGC_Approx_qPos_App} yields
\begin{align}
   \label{eq_CGC_Gen_App_q_neq_0} 
   \braket{s_\alpha,  m}{s_\alpha, m; k, 0} 
   = 1 + O\!\left( \frac{s_\alpha - m}{s_\alpha} \right) + \, \dots
\end{align}
The $\ldots$ consists of terms that contain additional powers of $(s_\alpha - m)/s_\alpha \, .$
Taylor-expanding the Clebsch--Gordan coefficient about 
$m = s_\alpha = M \, ,$ which is $O(\Sites) \, ,$ yields
\begin{align}
   \label{eq_CG_Expand_q0_App}
   \braket{s_\alpha, m}{s_\alpha, m; k, 0}
   & = 1 + O\!\left( \frac{m - M}{\Sites} \right)
   + O\!\left( \frac{s_\alpha - M}{\Sites} \right)
   + \ldots
\end{align}
A general term in the expansion is of
$O\! \left( [m - M]^B [s_\alpha - M]^C / \Sites^{B+C} \right)$, 
wherein $B, C \in \mathbb{Z}_{\geq 0}$.

\vspace{1em}
\emph{Taylor expansion of $\mathcal{T}^\KParen$:}
By assumption [see the text immediately above the non-Abelian ETH~\eqref{eq_nonAbelian_ETH}], 
$\mathcal{T}^\KParen (\mathcal{E}, \mathcal{S})$ is a smooth function of $\mathcal{E}/\Sites$ and $\mathcal{S}/\Sites$:
\begin{align}
   \frac{\partial^A}{\partial \mathcal{E}^A}
   \frac{\partial^C}{\partial \mathcal{S}^C} \,
   \mathcal{T}^\KParen(\mathcal{E}, \mathcal{S}) \bigg|_{\mathcal{E}=E, \mathcal{S}=M}
   = O\! \left( \frac{1}{\Sites^{A+C} } \right) \, .
\end{align}
Hence the Taylor expansion of $\mathcal{T}^\KParen (E_\alpha, s_\alpha)$ about $(E_\alpha = E, \, s_\alpha = M)$ has the form
\begin{align}
   \label{eq_Tk_Expand_App}
   \mathcal{T}^\KParen (E_\alpha, s_\alpha)
   & = \mathcal{T}^\KParen (E, M) 
   + O\! \left( \frac{E_\alpha - E}{\Sites} \right)
   + O\! \left( \frac{s_\alpha - M}{\Sites} \right)
   + \ldots
\end{align}
A general term in the expansion is of
$O\! \left( [E_\alpha - E]^A [s_\alpha - M]^C / \Sites^{A+C} \right)$, wherein $A, C \in \mathbb{Z}_{\geq 0}$.

\vspace{1em}
\emph{Combining the two Taylor expansions:}
We substitute the Taylor series~\eqref{eq_CG_Expand_q0_App} and~\eqref{eq_Tk_Expand_App} into the average~\eqref{eq_p_Avg_q0_a}, then multiply out.
As discussed in the main text's setup section, 
$\mathcal{T}^\KParen(E, M) = O(1) \, .$ Therefore,
\begin{align}
   \expval{ T^\KParen_0 }_p
   & = \sum_{\alpha, m}  p_{\alpha, m}
   \left[ \mathcal{T}^\KParen (E, M) 
   + O\! \left( \frac{E_\alpha - E}{\Sites} \right)
   + O\! \left( \frac{m - M}{\Sites} \right)
   + O\! \left( \frac{s_\alpha - M}{\Sites} \right)
   + \ldots \right] \, .
\end{align}
A general term has the form~$O\! \left( [E_\alpha - E]^A [m - M]^B [s_\alpha - M]^C / \Sites^{A+B+C} \right) \, .$
In the leading term, $\mathcal{T}^\KParen (E, M)$ can be factored out of the sum, which then equals one, by the normalization of 
$\{ p_{\alpha, m} \} \, .$
The general remaining term averages, by the moment condition~\eqref{eq_Moment_Condn}, to
$\leq O ( \Sites^{A+B+C-1} / \Sites^{A+B+C} )
= O(\Sites^{-1}) \, .$ Hence
\begin{align}
    \expval{ T^\KParen_0 }_p
    = \mathcal{T}^\KParen (E, M) + O(\Sites^{-1}) \, ,
\end{align}
as quoted in Eq.~\eqref{eq_p_Avg_q0_b}.

\section{Calculation of time average when 
\texorpdfstring{$M = O(\Sites)$}{M = O(\Sites)} and
\texorpdfstring{$q \neq 0$}{q \neq 0}}
\label{app_Time_Avg_q_neq_0}

Let us upper-bound the time-averaged expectation value~\eqref{eq_Time_Avg_NATS},
assuming that $M = O(\Sites)$ and $q \neq 0$:
\begin{align}
   \label{eq_Time_Avg_NATS_app}
   \overline{ \expval{ T^\KParen_{q \neq 0} }_t }
   = \sum_{\alpha, m} 
   C^*_{\alpha, m + q}  C_{\alpha, m}
   \braket{s_\alpha, m + q}{s_\alpha, m; k, q} \,
   \mathcal{T}^\KParen (E_\alpha, s_\alpha) \, .
\end{align}
When $q$ vanished (Suppl.~Note~\ref{app_q0}), 
we simplified the average using properties of 
$| C_{\alpha, m} |^2$.
We can achieve some of that simplification here, 
using the Cauchy--Schwarz inequality.
Once we do, the problem splits into three classes, associated with different $q$ values.
We upper-bound the time average~\eqref{eq_Time_Avg_NATS_app} for the classes sequentially.

\vspace{1em}
\emph{Application of Cauchy--Schwarz inequality:}
Define the vectors
$\vec{u}$ and $\vec{v}$ in terms of the components 
$u_{\alpha, m}^*
= C_{\alpha, m + q}^* \sqrt{\braket{s_\alpha, m + q}{s_\alpha, m; k, q} \mathcal{T}^\KParen (E_\alpha, s_\alpha)}$  and 
$v_{\alpha, m}
= C_{\alpha, m} 
\sqrt{\braket{s_\alpha, m + q}{s_\alpha, m; k, q} \mathcal{T}^\KParen (E_\alpha, s_\alpha)} \, .$
Any branch-cut convention can be applied to the square root.
Define the inner product 
$\vec{u} \cdot \vec{v} 
\coloneqq \sum_{\alpha, m} 
u_{\alpha, m}^* v_{\alpha, m} \, .$ 
The Cauchy--Schwarz inequality states that
$| \vec{u} \cdot \vec{v} | \leq \sqrt{\vec{u} \cdot \vec{u}} \, \sqrt{\vec{v} \cdot \vec{v}} \leq \max\{ \vec{u} \cdot \vec{u}, \, \vec{v} \cdot \vec{v} \} \, ,$ so
\begin{align}
   \label{eq_Time_Avg_q_neq_0_App1}
   \left| \overline{ \expval{ T^\KParen_{q \neq 0} }_t } \right|
   \leq \max\Bigg\{
   & \sum_{\alpha, m} \,
   \lvert C_{\alpha, m} \rvert^2 
   \big\lvert \braket{s_{\alpha}, m}{s_{\alpha}, m-q; k, q}
   \mathcal{T}^\KParen (E_\alpha, s_\alpha) \big\rvert \ ,
   \notag \\*
   & \sum_{\alpha, m} \,
   \lvert C_{\alpha, m } \rvert^2
   \big\lvert \braket{s_{\alpha}, m + q}{s_{\alpha}, m; k, q}
   \mathcal{T}^\KParen (E_\alpha, s_\alpha) \big\rvert
   \Bigg\} \, .
\end{align}
We have redefined $m+q \mapsto m$ in the first sum.

The sums are dominated by terms in which
\begin{align}
   \label{eq_Dominant_Terms}
   (E_\alpha \sim E, \, m \sim M, \, s_\alpha \sim M)
   \quad \text{and so, by the moment condition~\eqref{eq_Moment_Condn},} \quad
   s_\alpha - m \ll s_\alpha  \, ,
\end{align}
as when $q = 0$ (Suppl.~Note~\ref{app_q0}).
We approximate the Clebsch--Gordan coefficients 
under these conditions in Suppl.~Note~\ref{app_Approx_CGs}.
The result [Eqs.~\eqref{eq_CGC_Approx_qPos_App} and \eqref{eq_CGC_Approx_qNeg_App}] is
\begin{align}
   \label{eq_CGC_Gen_App_q_neq_0_b} 
   \braket{s_\alpha,  m + q}{s_\alpha, m; k, q} 
   = O\! \left( \left[
   \frac{s_\alpha - m}{s_\alpha} 
   \right]^{|q|/2} \right) + \, \dots
\end{align}
The $\dots$ consists of terms that contain additional powers of $(s_\alpha - m)/s_\alpha$ or $1/s_\alpha$.
The same asymptotic expansion characterizes the
$\braket{s_\alpha,  m}{s_\alpha, m-q; k, q}$
in Eq.~\eqref{eq_Time_Avg_q_neq_0_App1}.

We substitute the expansion~\eqref{eq_CGC_Gen_App_q_neq_0_b} into Ineq.~\eqref{eq_Time_Avg_q_neq_0_App1}.
Since $\mathcal{T}^\KParen (E_\alpha, s_\alpha) \leq O(1) \, ,$
\begin{align}
   \label{eq_CGC_bound_q2_App}
   \left| \overline{ \expval{ T^\KParen_{q \neq 0} }_t } \right|
   \leq \sum_{\alpha, m} \,
   \lvert C_{\alpha, m} \rvert^2
   \left\{ O\! \left( \left[
   \frac{s_\alpha - m}{s_\alpha} 
   \right]^{|q|/2} \right) 
   + \, \dots \right\} \, .
\end{align}
The $\dots$ consists of terms that contain additional powers of $(s_\alpha - m)/s_\alpha$ or $1/s_\alpha$.
We evaluate the bound for 
$|q| \geq 2 \, ,$ $q = 1 \, ,$ and $q = -1$ sequentially.

\vspace{1em}
\emph{Bounding the time-averaged expectation value when $|q| \geq 2$:} The moment condition~\eqref{eq_Moment_Condn}, 
together with $M = O(N)$, implies that 
$\langle (s_\alpha - m)^n/s_\alpha^n \rangle_\diag 
\leq O(N^{-1})$ for 
$n \geq 1$ [Eq.~\eqref{eq_sm_Moment_App}]. Therefore, since the $C_{\alpha, m}$'s are normalized to one, 
Ineq.~\eqref{eq_CGC_bound_q2_App} implies that
\begin{align}
   \left| \overline{ \expval{ T^\KParen_q }_t } \right|
   \leq O (\Sites^{-1}) \quad \text{if} \quad |q| \geq 2 \, .
\end{align} 
The time average equals the thermal average (zero),
to within $O(\Sites^{-1})$ corrections.

\vspace{1em}
\emph{Bounding the time-averaged expectation value when $q = + 1$:}
We return to the bound~\eqref{eq_CGC_bound_q2_App}.
Since $q = 1 \, ,$ the leading term averages to
$\langle [(s_\alpha - m)/s_\alpha]^{1/2} \rangle_\diag 
= O(N^{-1/2})$ in Eq.~\eqref{eq_Time_Avg_q_neq_0_App1}.
Therefore, we cannot immediately conclude that 
the time average $\leq O(\Sites^{-1}) \, .$

To demonstrate the time average's smallness, 
we return to Eq.~\eqref{eq_Time_Avg_NATS_app}.
We expect the same terms to dominate as when $|q| \geq 2$
[Eq.~\eqref{eq_Dominant_Terms}].
Accordingly, Eq.~\eqref{eq_CGC_Approx_qPos_App} approximates
the Clebsch--Gordan coefficient:
\begin{align}
    \braket{s_\alpha, m+1}{s_\alpha, m; k, 1}
   = - \sqrt{\frac{k(k+1)}{2}} \,
   \sqrt{\frac{s_\alpha - m}{s_\alpha}}
   \left[ 1 + O\! \left( \frac{s_\alpha - m}{s_\alpha} \right) + \dots \right] \, .
\end{align}
Substituting into Eq.~\eqref{eq_Time_Avg_NATS_app} yields
\begin{align}
   \label{eq_Time_Avg_q1_a}
   \overline{ \expval{ T^\KParen_{q=1} }_t }
   & = - \sqrt{\frac{k(k+1)}{2}}
   \sum_{\alpha, m} 
   C_{\alpha, m+1}^* C_{\alpha, m}
   \sqrt{ \frac{s_\alpha - m}{s_\alpha} } \,
   \left[ 1 + O\! \left( \frac{s_\alpha - m}{s_\alpha} \right) + \dots \right]
   \mathcal{T}^\KParen (E_\alpha, s_\alpha) \, .
\end{align}
We can Taylor-expand everything except 
the $\sqrt{s_\alpha - m}$ about
$(E_\alpha = E, \, m = M, \, s_\alpha = M) \, .$
Since $\mathcal{T}^\KParen (E_\alpha, s_\alpha)
= \mathcal{T}^\KParen (E, M)
+ O\! \left( \frac{E_\alpha - E}{\Sites} \right)
+ O\! \left( \frac{s_\alpha - M}{\Sites} \right) + \dots$
[Eq.~\eqref{eq_Tk_Expand_App}],
\begin{align}
   \label{eq_Time_Avg_q1_b}
   \overline{ \expval{ T^\KParen_{q=1} }_t }
   = -\sqrt{ \frac{k (k + 1)}{2} } \,
   \sum_{\alpha, m} 
   & \, C_{\alpha, m+1}^* C_{\alpha, m}
   \sqrt{\frac{s_\alpha - m}{M}} \notag \\*
   &\times 
   \left[ \mathcal{T}^\KParen (E, M)
   + O\! \left( \frac{E_\alpha - E}{\Sites} \right)
   + O\! \left( \frac{m - M}{\Sites} \right)
   + O\! \left( \frac{s_\alpha - M}{\Sites} \right) 
   + \dots \right] \, .
\end{align}
The leading term, involving $\mathcal{T}^\KParen (E, M)$, is the problematic one [for proving that the time average is 
$\leq O(\Sites^{-1})$]: 
The leading term looks to be of $O(\Sites^{-1/2})$.
However, this term is actually proportional to
$\langle S_+ \rangle \equiv \expval{S_x} + i \expval{S_y}$,
which vanishes by assumption
[the sentence immediately below Eq.~\eqref{eq_M_def}, in the main text's setup].
To prove the proportionality, we substitute 
the initial-state expansion 
$\ket{\psi(0)} = \sum_{\alpha,m} C_{\alpha,m} \ket{\alpha,m}$ 
and the raising-operator equation
$S_+ \ket{\alpha, m}
= \sqrt{(s_\alpha - m)(s_\alpha + m + 1)} \, \ket{\alpha,m+1}$
into the expectation value
$\expval{ S_+ }
= \sum_{\alpha, m}
C_{\alpha, m + 1}^* C_{\alpha, m}
\sqrt{(s_\alpha - m)(s_\alpha + m + 1)} \, .$
Taylor-expanding
$\sqrt{s_\alpha + m + 1}
= \sqrt{2M + 1} \, 
\left[1 + O\! \left( \frac{s_\alpha - M}{\Sites} \right)
+ O\! \left( \frac{m - M}{\Sites}  \right) + \dots \right]$
yields
\begin{align}
   \label{eq_SPlus}
   \expval{ S_+ }
   = \sqrt{2M + 1}
   \sum_{\alpha, m}
   C_{\alpha, m + 1}^* C_{\alpha, m}
   \sqrt{s_\alpha - m} \,
   \left[ 1 + O\! \left( \frac{s_\alpha - M}{\Sites} \right)
              + O\! \left( \frac{m - M}{\Sites} \right) + \dots
   \right] = 0 \, .
\end{align}
The leading-order term is proportional to
the leading-order term in Eq.~\eqref{eq_Time_Avg_q1_b}. 
Therefore, at leading order,
$\overline{ \expval{ T^\KParen_{q=1} }_t } = 0 \, .$
The higher-order terms in Eq.~\eqref{eq_Time_Avg_q1_b} can be shown to evaluate to $\leq O(\Sites^{-1})$; one repeats the Cauchy--Schwarz argument used for $|q| \geq 2$.
Therefore,
\begin{align}
   \left| \overline{ \expval{ T^\KParen_{q = 1} }_t } \right|
   \leq O (\Sites^{-1}) \, .
\end{align}

\vspace{1em}
\emph{Bounding the time-averaged expectation value when $q = -1$:}
The proof is almost the same as for $q = 1 \, .$
Instead of $\langle S_+ \rangle = 0$, we use 
$\langle S_- \rangle \equiv \expval{S_x} - i \expval{S_y} = 0$:
\begin{align}
   \left| \overline{ \expval{ T^\KParen_{q = -1} }_t } \right|
   \leq O (\Sites^{-1}) \, .
\end{align}
Thus, for all $q \neq 0 \, ,$ 
the time average equals the thermal average (zero),
to within $O(\Sites^{-1})$ corrections.

\section{Approximation of Clebsch--Gordan coefficients}
\label{app_Approx_CGs}

Here, we approximate the Clebsch--Gordan coefficients $\braket{s_\alpha,  m + q}{s_\alpha, m; k, q}$ when 
$s_\alpha \gg 1$ and $s_\alpha - m \ll s_\alpha$.
As throughout this paper, $k, q = O(1)$.

The general expression for Clebsch--Gordan coefficients
is~\cite[Eq.~(2.41)]{Bohm_Quantum_Book}
\begin{align}
   \label{eq_CGC_ApproxApp} &
   \braket{s, m}{s', m'; k, q}
   = \delta_{m, \, m' + q} 
   \\ \nonumber & 
   \times \sqrt{ \frac{
   (2 s + 1) \, (s + s' - k)! \, 
   (s - s' + k)! \, (s' + k - s)! \, 
   (s + m)! \, (s - m)! \, 
   (s' - m')! \, (s' + m')! \, 
   (k - q)! \, (k + q)!}{
   (s + s' + k + 1)!} }
   \\ \nonumber & \times
   \sum_\ell \frac{ (-1)^\ell }{
   \ell! \, (s' + k - s - \ell)! \, 
   (s' - m' - \ell)! \, (k + q - \ell)! \, 
   (s - k + m' + \ell)! \, 
   (s - s' - q + \ell)!} \, .
\end{align}
The final line's sum runs over all integer $\ell$ values 
for which every factorial's argument is nonnegative.
This expression holds for $m \geq 0$ and $s' > k$.
We set $s' = s$, as in 
the time-averaged expectation value~\eqref{eq_Time_Avg_NATS}.
We also set $m = m'+q$ and drop the primes:
\begin{align}
   \label{eq_Approx_CG_App_1}
   \braket{s, m+q}{s, m; k, q}
   & = 
   \sqrt{\frac{(2s + 1) \, (2 s - k)! \,
   (k!)^2 \,
   (s + m + q)! \, (s - m - q)! \,
   (s - m)! \, (s + m)! \,
   (k - q)! \, (k + q)!
   }{ (2 s + k + 1)! }}
   \nonumber \\ & \quad \times
   \sum_\ell \frac{ (-1)^\ell }{
   \ell! \, (k - \ell)! \, (s - m - \ell)! \,
   (k + q - \ell)! \, (s - k + m + \ell)! \,
   (\ell - q)! } \, .
\end{align}
This expression holds for $m + q \geq 0$ and $s > k$.
Both conditions are satisfied in the regime of interest, wherein $s \gg 1$ and $s-m \ll s$, while $k,q = O(1)$.
The sum runs over the integers $\ell$ for which 
each factorial's argument is nonnegative.

Together, the factorials imply three upper bounds 
and three lower bounds on $\ell$:
\begin{align}
   \label{eq_ell_Bd1}
   & \ell  \leq k \, , \\
   \label{eq_ell_Bd2}
   & \ell  \leq  s - m \, , \\
   \label{eq_ell_Bd3}
   & \ell  \leq  k + q \, , \\
   \label{eq_ell_Bd4}
   & \ell \geq k - s - m  \, , \\
   \label{eq_ell_Bd5}
   & \ell \geq q \, ,
   \quad \text{and} \\
   \label{eq_ell_Bd6}
   & \ell \geq 0 \, .
\end{align}
We initially assume that $q \geq 0$, then generalize.
Consequently, Ineq.~\eqref{eq_ell_Bd1} subsumes Ineq.~\eqref{eq_ell_Bd3}, and
Ineq.~\eqref{eq_ell_Bd5} subsumes Ineq.~\eqref{eq_ell_Bd6}.
As $s \gg 1$ and $s-m \ll s$, Ineq.~\eqref{eq_ell_Bd4} encodes a trivially negative lower bound.
The constraints on $\ell$ reduce to
\begin{align}
   \label{eq_Ell_Bd}
   \ell \in \big\{ q, q+1, \ldots, 
   \min \{k, s - m \} \big\} \, .
\end{align}

Let us extract the asymptotics of the Clebsch--Gordan coefficient in the limit as $s \to \infty$, 
assuming $(s-m)/s \to 0$.
For convenience, we change variables from $m$ to
$\Delta \coloneqq s - m$.
Grouping the $s$-dependent factors together yields
\begin{align}
   \label{eq_Approx_CG_App_2}
   \braket{s, m+q}{s, m; k, q}
   & = \sum_\ell 
   \frac{ (-1)^\ell \,  k! \, \sqrt{ (\Delta - q)! \, 
          \Delta! \, (k - q)! \, (k + q)! }}{
          \ell! \, (k - \ell)! \, (\Delta - \ell)! \,
          (k + q - \ell)! \, (\ell - q)! }
   \nonumber \\ & \qquad \times
   \sqrt{\frac{ (2s + 1) (2 s - k)!
   (2 s - \Delta + q)! \,
   (2 s - \Delta)! }{
   (2 s + k + 1)! }} \:
   \frac{1}{(2 s - \Delta - k + \ell)!} \, .
\end{align}
The $s$-dependent factorials are all large in the limit of interest, so we approximate them using Stirling's formula:
\begin{align}
   x! = \exp\!\left( x \ln x - x
   + \frac{1}{2} \ln (2 \pi x)
   + O(1 / x) + \dots \right) \, .
\end{align}
We take the $s$-dependent expression's natural log,
expand in powers of $1 / s$, and exponentiate.
The $s$-dependent factor is
\begin{align}
   \label{eq_s_Factor}
   (2 s)^{q/2 - \ell} \,
   \exp\Big( O(\Delta/s) + \cdots \Big)
   = (2 s)^{q/2 - \ell} \,
   \Big[ 1 + O(\Delta/s) + \cdots 
   \Big] \, .
\end{align}
Therefore, the least possible $\ell$ value dominates 
the $\sum_\ell$ in Eq.~\eqref{eq_Approx_CG_App_2}.
By~\eqref{eq_Ell_Bd}, that $\ell$ value is $q$.

Let us approximate the $\sum_\ell$ with the $\ell = q$ term,
while replacing the $s$-dependent factor with~\eqref{eq_s_Factor}.
We revert notation from $\Delta$ to 
$s - m$. The result is
\begin{align}
   \label{eq_CGC_Approx_qPos_App}
   \braket{s, m+q}{s, m; k, q}
   = \frac{(-1)^q}{q! (2 s)^{q/2} } \,
   \left( \frac{
   (s - m)! \,  (k + q)!}{
   (s - m - q)! \, (k - q)!}
   \right)^{1/2}
   \left[ 1 + O\! \left( \frac{s - m}{s} \right) + \dots
   \right] \, , \qquad q \geq 0 \, .
\end{align}

Now, suppose that $q < 0$.
The bounds~\eqref{eq_Ell_Bd} on $\ell$ become
$\ell \in \big\{ 0, 1, \ldots, 
 \min \{k-q, s - m \} \big\} \, .$
In Eq.~\eqref{eq_Approx_CG_App_2},
$\ell = 0$ labels the sum's dominant term.
The $s$-dependent factor approximates to
$(2 s)^{q/2} 
[1 + O\! \left( \frac{s - m}{s} \right) + \dots ] \, .$
Again, we substitute into and approximate Eq.~\eqref{eq_Approx_CG_App_2}.
The result is
\begin{align}
   \label{eq_CGC_Approx_qNeg_App}
   \braket{s, m+q}{s, m; k, q}
   = \frac{1}{|q|! (2 s)^{|q|/2} } \,
   \left( \frac{
   (s - m + |q|)! \,  (k + |q|)!}{
   (s - m)! \, (k - |q|)!}
   \right)^{1/2}
   \left[ 1 + O\! \left( \frac{s - m}{s} \right) + \dots
   \right] \, , \qquad q < 0 \, .
\end{align}

\section{Taylor expansion of 
\texorpdfstring{$\mathcal{T}^\KParen(\mathcal{E}, \mathcal{S})$}{\mathcal{T}^\KParen(\mathcal{E}, \mathcal{S})}
about
\texorpdfstring{$\mathcal{S} = 0$}{\mathcal{S} = 0}}
\label{app_Form_of_Tk}

The main text shows that, if the total magnetization $M=0$,
noncommuting charges can lead to anomalous thermalization. 
Our arguments rely on a claim about 
the smooth function $\mathcal{T}^\KParen (\mathcal{E}, \mathcal{S})$:
Suppose that $k \geq 0$ is even and 
$\mathcal{E} = O(\Sites) \, .$
Consider Taylor-expanding 
$\mathcal{T}^\KParen (\mathcal{E}, \mathcal{S})$
about $\mathcal{S} = 0$.
For some Hamiltonians $H$ and operators $T^\KParen_q \, ,$ 
the Taylor expansion can have a nonvanishing
$O(\mathcal{S} / \Sites)$ term:
\begin{align}
   \label{eq_Form_of_Tk_app}
   \mathcal{T}^\KParen (\mathcal{E}, \mathcal{S})
   = \mathcal{T}^\KParen (\mathcal{E}, 0) + O(\mathcal{S} / \Sites) + \ldots .
\end{align}
Furthermore, the $O(1)$ term $\mathcal{T}^\KParen (\mathcal{E}, 0)$ vanishes whenever $k > 0$.
Here, we argue for the claim.

\vspace{1em}
\emph{Argument that $\mathcal{T}^\KParen (\mathcal{E}, 0) = 0$ when $k > 0$:}
When $s_\alpha = 0$, the system lacks spin angular momentum and so any preferred direction.
Therefore, all rotationally noninvariant operators' expectation values must vanish.
$T^\KParen_q$ is rotationally noninvariant for all $q$, because $k > 0$.
To identify the implications for $\mathcal{T}^\KParen (\mathcal{E}, 0) \, ,$
we evaluate 
$\bra{\alpha, m}  T^\KParen_q  \ket{\alpha, m}$
(which vanishes)
on an $s_\alpha = 0$ eigenstate $\ket{\alpha, m} \, .$
We invoke the Wigner--Eckart theorem~\eqref{eq_WigEck} and
the non-Abelian ETH~\eqref{eq_nonAbelian_ETH}.
The associated Clebsch--Gordan coefficient equals one.
Hence $\bra{\alpha, m}  T^\KParen_q  \ket{\alpha, m}
= \mathcal{T}^\KParen (E_\alpha, 0)\, .$ 
The LHS vanishes, as argued above.
Hence the $O(1)$ term $\mathcal{T}^\KParen (\mathcal{E}, 0) = 0 \, .$

\vspace{1em}
\emph{Argument for the $O(\mathcal{S}/\Sites)$ term in Eq.~\eqref{eq_Form_of_Tk_app}:}
No deductive arguments preclude the form~\eqref{eq_Form_of_Tk_app}, to our knowledge.
Hence there is no reason to believe that 
the $O(\mathcal{S}/\Sites)$ term is absent.
Beyond this ``everything not forbidden is compulsory'' reasoning, we also argue for 
the $O(\mathcal{S}/\Sites)$ term's plausibility.
Bound states underlie the argument.
To provide intuition, we first address a more familiar setting, in which just particles are conserved. We then replace the particles with spin quanta.

\vspace{1em}
\emph{Bound particles:}
Imagine a lattice in which only the global particle-number operator, 
$\mathcal{N} \, ,$ is conserved. 
Consider a global state $\ket{\psi}$ of 
uniformly distributed two-particle bound states.
Wherever a particle appears, another particle appears beside it.
Denote by $\mathcal{N}_j$ 
the site-$j$ particle-number operator.
For an arbitrary $j$, we estimate the correlator
$\bra{\psi} \mathcal{N}_j \mathcal{N}_{j+1} \ket{\psi}$.

The correlator equals the joint probability 
$p$(particle at site $j$, particle at site $j+1$). 
Semiclassically, this joint probability equals
$p$(particle at site $j+1 | $particle at site $j$) 
$\times$ 
$p$(particle at site $j$).
The latter probability equals 
$O (\bra{\psi} \mathcal{N} \ket{\psi} / \Sites) \, ,$
by the state's uniformity.
The conditional probability is $O(1)$, 
because the particles are bound.
Hence the joint probability
$p$(particle at site $j$, particle at site $j+1$)
$= O (\bra{\psi} \mathcal{N} \ket{\psi} / \Sites) \, .$

\vspace{1em}
\emph{Bound quanta of spin:}
We reason about noncommuting charges by 
analogy with the preceding argument.
First, suppose that $k \geq 2$.
Let the Hamiltonian have a finite-energy-density eigenstate 
$\ket{\alpha, m {=} s_\alpha}$
that contains bound clusters of $k$ $z$-type charges.
For example, ferromagnetic couplings can cause
neighboring spins to point in the same direction.
If that direction is $\hat{z}$,
the state contains bound $z$-charges.
If $H$ has some degree of uniformity,
so can $\ket{\alpha, s_\alpha}$. 
The amount of charge in the global system is essentially $s_\alpha$.

A local operator of interest has the form~\footnote{
To produce a $T^\KParen_0$ operator, one would begin with the right-hand side (RHS) of~\eqref{eq_Approx_TKq_App}, multiply it by a constant, and subtract off terms to make the expression orthogonal to all the lower-$k$ operators. For example, the main text features a quadrupole operator. This $T^\2_0$ results from subtracting the $T^\0_0$ operator $\vec{s}_i \cdot \vec{s}_j$ from $3 s_{i,z} s_{j,z}$. However, the RHS of~\eqref{eq_Approx_TKq_App} forms the essence of $T^\KParen_0$.
}
\begin{align}
   \label{eq_Approx_TKq_App}
   T^\KParen_0
   \sim s_{j_1,z} \, s_{j_2,z} \ldots s_{j_k,z} \, .
\end{align}
Recall that $s_{j,z}$ denotes qubit $j$'s $z$-type spin operator.
This $T^\KParen_0$ has an expectation value,
in a joint eigenstate $\ket{\alpha, m {=} s_\alpha}$, 
that is essentially a $k$-point correlator:
\begin{align}
   \label{eq_LinearTerm_Help1}
   \bra{\alpha, s_\alpha} T^\KParen_0 \ket{\alpha, s_\alpha}
   \sim \bra{\alpha, s_\alpha} 
   s_{j_1,z} \, s_{j_2,z} \ldots s_{j_k,z}
   \ket{\alpha, s_\alpha} \, .
\end{align}

Similarly to in the particle-number example, this correlator is essentially the joint probability
\begin{align}
   \label{eq_LinearTerm_Help2}
   & p( \text{site $j_1$ contains a quantum of $z$-type charge,
site $j_2$ contains a quantum of $z$-charge,} \ldots, 
   \nonumber \\ & \quad
   \text{site $j_k$ contains a quantum of $z$-charge} ) \, .
\end{align}
Semiclassically, this joint probability equals
\begin{align}
   \label{eq_LinearTerm_Help3} &
   p( \text{site $j_2$ contains a quantum of $z$-charge,
        \ldots,
        site $j_k$ contains a quantum of $z$-charge}
    \nonumber \\ & \quad
    | \text{site $j_1$ contains a quantum of $z$-charge} ) \times
   p( \text{site $j_1$ contains a quantum of $z$-charge} ) \, .
\end{align}

The final probability is $O(s_\alpha / \Sites) \, ,$
by the state's uniformity.
The conditional probability is $O(1) \, ,$
if $j_1$ lies close to the other $j$'s,
because the charges form bound clusters.
Hence the joint probability is $O(s_\alpha / \Sites) \, .$
So, semiclassically,
$\bra{\alpha, s_\alpha} T^\KParen_0 \ket{\alpha, s_\alpha}
= O(s_\alpha / \Sites) \, .$
The LHS is essentially 
$\mathcal{T}^\KParen (E_\alpha, s_\alpha) \, ,$
by the Wigner--Eckart theorem~\eqref{eq_WigEck}
and the non-Abelian ETH~\eqref{eq_nonAbelian_ETH}.
Hence $\mathcal{T}^\KParen (E_\alpha, s_\alpha)
= O(s_\alpha / \Sites) \, .$

If $k=0$, a similar argument concerns a local operator $T^\0_0 = -\vec{s}_j \cdot \vec{s}_{j'}$.
Start with a ``resonating valence bond'' (RVB) state,
a superposition of local dimer coverings (local pairings of the qubits into singlets). This state has zero total spin ($s_\alpha = 0$).
The expectation value $\big\langle T^\0_0 \big\rangle 
= \expval{- \vec{s}_j \cdot \vec{s}_{j'}}$ quantifies
the probability that qubits $j$ and $j'$ form a singlet.
This probability is $O(1)$ in the RVB state, 
if $j$ lies close to $j'$. 
Imagine that a few of the singlets are broken---a small fraction $\rho$.
Furthermore, the broken singlets are distributed uniformly throughout the system. 
The resulting state has nonzero but small total spin:
$s_\alpha/\Sites \sim \rho$. The expectation value 
$\big\langle T^\0_0 \big\rangle 
= \expval{-\vec{s}_j \cdot \vec{s}_{j'}} 
= O(1) + O(s_\alpha/\Sites) \, .$ 
The second term equals the probability that the $j$-and-$j'$ singlet is broken.
Hence, by the same logic as for $k \geq 2$,
$\mathcal{T}^\0 (E_\alpha, s_\alpha)
= O(1) + O(s_\alpha / \Sites) \, .$

\vspace{2em}
\section{Opportunity for anomalous thermalization of \\
rotationally noninvariant operators 
\texorpdfstring{$T^{(k \geq 2)}_q$}{T^{(k \geq 2)}_q} 
when \texorpdfstring{$M = 0$}{M = 0}}
\label{app_Anomaly2}

The main text illustrates potential anomalous thermalization with a rotationally noninvariant operator $T^{(k>0)}_q$.
This supplementary note covers the topic in greater depth.
To illustrate the generality of such potential anomalous thermalization, we focus on operators 
$T^{(k \geq 2)}_{q=1}$,
rather than the main text's $T^\2_0$, here.

The strategy is as follows.
We construct a time-independent state 
$\ket{\psi(t)} = \ket{\psi(0)} \; \forall t$ 
that has the properties stipulated in our setup, 
as well as $M = 0$.
Then, we focus on operators $T^\KParen_{q=1}$,
for an even $k \geq 2 \, .$
The time average 
$\overline{ \langle T^\KParen_1 \rangle_t } \, ,$
we show, differs from the thermal average 
$\langle T^\KParen_1 \rangle_\th$ at the anomalously large $O(\Sites^{-1/2}) \, .$
The argument relies on the smooth function
$\mathcal{T}^\KParen(E_\alpha, s_\alpha)$'s 
having a term $\sim s_\alpha / \Sites$
in its Taylor expansion about $s_\alpha = 0$
(Suppl.~Note~\ref{app_Form_of_Tk}).

Consider an initial state $\ket{\psi(0)}$ in
a Hamiltonian eigenspace labeled by $\alpha = A \, .$ 
Let the eigenenergy $E_A = O(\Sites)$
and spin quantum number $s_A = O(\Sites^{1/2}) \, .$
For some to-be-specified magnetic spin quantum number
$\bar{m} \in [-s_A, \, s_A] \, ,$
\begin{align}
   \label{eq_Init_State_Anom2}
   \ket{\psi(0)}
   \coloneqq \frac{1}{2} ( \ket{A, \bar{m}} + \ket{A, \bar{m} + 1 }
   + \ket{A, -\bar{m} } - \ket{A, - \bar{m} - 1} ) \, .
\end{align}
One can check directly that 
$M = \langle S_z \rangle = 0$ and
$\var (S_z) = \frac{1}{2} [\bar{m}^2 + (\bar{m}+1)^2] = O(\Sites) \, .$
That $\langle S_{x,y} \rangle = 0$ follows from
(i) the decompositions of $S_x$ and $S_y$ in terms of $S_\pm$
and (ii) the ladder operators' actions on an $S_z$ eigenstate,
$S_\pm \ket{A, m}
= \sqrt{s_A (s_A + 1) - m (m \pm 1)} \:
\ket{A, m \pm 1} \, .$
The same ingredients imply that
$\langle S_{x,y}^2 \rangle
= O(s_A^2) + O(\bar{m}^2)
= O(\Sites)$;
hence $\var(S_{x,y}) = O(\Sites) \, .$
The energy variance, $\var(H)$, vanishes by construction.
Hence $\ket{\psi(0)}$ has the properties stipulated in
the main text's setup section, including the variance conditions~\eqref{eq_Var_H}--\eqref{eq_Var_xy}.

Having prescribed an initial state,
we shift focus to an operator $T^\KParen_q \, .$
Let $q = 1 \, .$
To calculate the time-averaged expectation value,
we substitute the $C_{\alpha, m}$'s from Eq.~\eqref{eq_Init_State_Anom2} into Eq.~\eqref{eq_Time_Avg_NATS}:
\begin{align}
   \label{eq_Time_Avg_App_Anom2}
   \overline{ \expval{ T^\KParen_1 }_t }
   & = \frac{1}{4} \Big( 
   \braket{s_A, \bar{m} + 1}{s_A, \bar{m}; k, 1}
   - \braket{s_A, -\bar{m}}{s_A, - \bar{m} - 1; k, 1} \Big) \,
   \mathcal{T}^\KParen (E_A, s_A) \, .
\end{align}
The Clebsch--Gordan coefficients obey the symmetry relation
\begin{align}
   \label{eq_CG_Sym}
   \braket{s_\alpha, m + 1}{s_\alpha, m; k, 1}
   = (-1)^{k+1} \braket{s_\alpha, -m}{
                        s_\alpha - m - 1; k, 1} 
\end{align}
\cite[Eq.~(2.42)]{Bohm_Quantum_Book}.
Consequently, if $k$ is odd, the time average~\eqref{eq_Time_Avg_App_Anom2} vanishes.
Since the $S_\pm$ are $T^\1_{\pm 1}$ operators, we have corroborated 
our earlier conclusion that
$\langle S_{x,y} \rangle = 0 \, .$
Suppose, instead, that $k \geq 2$ is even.
The symmetry~\eqref{eq_CG_Sym} reduces 
the time average~\eqref{eq_Time_Avg_App_Anom2} to
\begin{align}
   \label{eq_Time_Avg_App_Anom2_2}
   \overline{ \expval{ T^\KParen_1 }_t }
   = \frac{1}{2}
   \braket{s_A, \bar{m}+1}{s_A, \bar{m}; k, 1} \,
   \mathcal{T}^\KParen (E_A, s_A) \, .
\end{align}

We now Taylor-expand the smooth function $\mathcal{T}^\KParen (E_A, s_A)$ about $s_A = 0 \, .$ 
Generally, the Taylor expansion can have the form
$\mathcal{T}^\KParen (E_A, s_A) = O(1) + O(s_A / \Sites) + \ldots \, .$
We argued in Suppl.~Note~\ref{app_Form_of_Tk} that the $O(1)$ term vanishes if $k > 0$, but that the $O(s_A / \Sites)$ term is nonzero under a physically reasonable assumption.
This term is $O( \Sites^{-1/2} )$, by our choice of $s_A = O(\Sites^{1/2})$.
Hence Eq.~\eqref{eq_Time_Avg_App_Anom2_2} has the form
\begin{align}
   \label{eq_Time_Avg_App_Anom2_3}
   \overline{ \expval{ T^\KParen_1 }_t }
   = \frac{1}{2}
   \braket{s_A, \bar{m}+1}{s_A, \bar{m}; k, 1} \times
   O( \Sites^{-1/2} ) \, .
\end{align}
We can choose $\bar{m}$ to make the Clebsch--Gordan coefficient be $O(1) \, .$
\{In fact, the Clebsch--Gordan coefficient is $O(1)$ for most choices of $\bar{m} \in [-s_A, s_A]$ that are not too close to $\pm s_A$.\}
Substituting into Eq.~\eqref{eq_Time_Avg_App_Anom2_3} yields
\begin{align}
   \overline{ \expval{ T^\KParen_1 }_t }
   = O ( \Sites^{-1/2} ) \, .
\end{align}
The thermal average $\langle T^\KParen_1 \rangle_\th$ vanishes,
because $q \neq 0$, by Eq.~\eqref{eq_Therm_ExpVal1}.
Hence the time average differs from the thermal average
at $O( \Sites^{-1/2} ) \, .$

\vspace{2em}
\section{Opportunity for anomalous thermalization of \\ rotationally invariant operators \texorpdfstring{$T^\0_0$}{T^\0_0}
when \texorpdfstring{$M = 0$}{M = 0}}
\label{app_T00_Avgs}

The main text illustrated potential anomalous thermalization with
a rotationally noninvariant operator $T^{(k>0)}_q$.
Here, we show that rotationally invariant operators 
$T^\0_0$, too, may thermalize anomalously if $M = 0$.

Consider an arbitrary $T^\0_0$.
The corresponding Clebsch--Gordan coefficients
$\braket{s_\alpha, m}{s_\alpha, m; 0, 0} = 1$.
Hence the thermal average~\eqref{eq_Therm_ExpVal1} and
time average~\eqref{eq_Time_Avg_NATS} reduce to
\begin{align}
   \label{eq_k0_Therm_Avg}
   & \big\langle T^\0_0 \big\rangle_\th
   = \frac{1}{Z} \sum_{\alpha, m} e^{-\beta E_\alpha} \,
   \mathcal{T}^\0 (E_\alpha, s_\alpha)
   \quad \text{and} \\ &
   \label{eq_k0_Time_Avg}
   \overline{ \big\langle T^\0_0 \big\rangle_t }
   = \sum_{\alpha, m}
   | C_{\alpha, m} |^2 \,
   \mathcal{T}^\0 (E_\alpha, s_\alpha) \, .
\end{align}
We outline the rest of the argument here,
then detail it in Suppl.~Note.~\ref{app_Calc_Therm_T00}.

In each of Eqs.~\eqref{eq_k0_Therm_Avg} and~\eqref{eq_k0_Time_Avg}, we Taylor-expand the smooth function
$\mathcal{T}^\0 (E_\alpha, s_\alpha)$ about 
$(E_{\alpha}=E, \, s_{\alpha} = 0)$, then average term by term.
In the expansion, a general term is a moment 
$\langle (E_{\alpha}-E)^A \, (s_{\alpha})^C \rangle$
times a corresponding derivative of $\mathcal{T}^\0$.
The leading ($A=C=0$) term averages to $\mathcal{T}^\0 (E, 0)$ in both~\eqref{eq_k0_Therm_Avg} and~\eqref{eq_k0_Time_Avg}:
The thermal and time averages equal each other to zeroth order.
Terms with $A \geq 1$ or $C \geq 2$ all average to 
$\leq O(\Sites^{-1})$, as when $M = O(\Sites)$.
The remaining $(A = 0, \, C=1)$ term is linear in the spin density, $s_{\alpha}/\Sites$.

This $O(s_\alpha/\Sites)$ term, we have argued, can be nonzero for suitable Hamiltonians and operators.
See the main text and Suppl.~Note~\ref{app_Form_of_Tk} 
for an argument based on bound states.
The $O(s_{\alpha}/\Sites)$ term, in the thermal average~\eqref{eq_k0_Therm_Avg}, 
evaluates to $O( \Sites^{-1/2})$
(Suppl.~Note~\ref{app_Calc_Therm_T00}).

In contrast, consider the time average~\eqref{eq_k0_Time_Avg}.
We can engineer $C_{\alpha, m}$ to be large only when 
$s_\alpha = O(1)$.
The $O(s_{\alpha} / \Sites)$ term will then time-average to $O(\Sites^{-1})$.
%
The time average will deviate from the thermal average by
\begin{align}
   \label{eq_Correction}
   \expval{ T^\0_0 }_\th
   - \overline{ \expval{ T^\0_0 }_t }
   = O ( \Sites^{-1/2} ) 
   > O ( \Sites^{-1} ) \, .
\end{align}

\subsection{Thermal average of rotationally invariant operators
at \texorpdfstring{$M = 0$}{M = 0}}
\label{app_Calc_Therm_T00}

Here, we evaluate in greater detail 
the thermal average~\eqref{eq_k0_Therm_Avg}.
The calculation proceeds as follows: 
Upon approximating the sum as an integral, we use Laplace's method (similar to the saddle-point approximation). We Taylor-expand the smooth $\mathcal{T}^\0 (E_\alpha, s_\alpha)$ near the peak, then integrate term by term.

The summand is a smooth function of 
$E_\alpha / \Sites$ and $s_\alpha / \Sites$ when $\Sites \gg 1 \, ,$
by assumption [see the text immediately above the non-Abelian ETH~\eqref{eq_nonAbelian_ETH}].
Therefore, we can replace the sum over states with an integral over energy and spin quantum numbers (treated as continuous variables), weighted by the density of states $e^{\ent}$:
\begin{align}
   \label{eq_k0_Therm_Int_App}	
   \expval{ T^\0_0 }_\th 
   \approx \frac{1}{ Z' } 
   \int_{E_\Min}^{E_\Max} d\mathcal{E} 
   \int_0^{s_\Max} d\mathcal{S}  \ e^{\ent(\mathcal{E},\mathcal{S}) 
	   - \beta \mathcal{E} } \, 
   \mathcal{T}^\0 (\mathcal{E},\mathcal{S}) \, .
\end{align}
The normalization condition $\langle \id \rangle_\th = 1$ fixes the effective partition function $Z' \, .$

We evaluate the integral using Laplace's method (the saddle-point approximation, but for real variables). For most fixed values of $\mathcal{E}/\Sites$ and $\mathcal{S}/\Sites \, ,$ the exponent $\ent(\mathcal{E},\mathcal{S}) - \beta \mathcal{E} 
= O(\Sites) \, ,$
so the integrand peaks 
steeply about this function's global maximum.
The exponent is stationary where its first derivatives vanish: 
\begin{align}
   \label{eq_Max_App}
   \frac{\partial \ent}{\partial \mathcal{E}} - \beta = 0 \, ,
   \quad \text{and} \quad
   \frac{\partial \ent}{\partial \mathcal{S}} = 0 .
\end{align}
The first condition is the usual thermodynamic definition of the inverse temperature $\beta$.
The thermodynamic entropy's concavity suggests that~\eqref{eq_Max_App} has a unique solution
$(\mathcal{E}, \mathcal{S}) = (\mathcal{E}_*, \mathcal{S}_*)$, 
at which the exponent attains its global maximum.
We Taylor-expand the exponent about this maximum:
\begin{align}
	\label{eq_Laplace_exp_App}
	\ent(\mathcal{E},\mathcal{S}) - \beta \mathcal{E}
	\approx \ent(\mathcal{E}_*,\mathcal{S}_*) - \beta \mathcal{E}_*
	+ \frac{1}{2} \left( \mathcal{E} - \mathcal{E}_*, \, \mathcal{S} - \mathcal{S}_* \right)^T
	\left[\nabla^2_\hess \ent \right]_* 
	\left( \mathcal{E} - \mathcal{E}_*, \, \mathcal{S} - \mathcal{S}_* \right) \, .
\end{align}
$\left[\nabla^2_\hess \ent \right]_*$ denotes the Hessian matrix of $\ent(\mathcal{E},\mathcal{S})$, evaluated at 
$(\mathcal{E}_*, \mathcal{S}_*) \, .$
Equation~\eqref{eq_k0_Therm_Int_App} reduces to
\begin{equation}
    \label{eq_k0_Therm_Int2_App}
	\expval{ T^\0_0 }_\th 
	\approx \frac{1}{ Z'' } 
	\int_{E_\Min}^{E_\Max} d\mathcal{E} 
	\int_0^{s_\Max} d\mathcal{S} \
	e^{- \frac{1}{2} \left( \mathcal{E} - \mathcal{E}_*, \, \mathcal{S} - \mathcal{S}_* \right)^T
	\left[-\nabla^2_\hess \ent \right]_*
	\left( \mathcal{E} - \mathcal{E}_*, \, \mathcal{S} - \mathcal{S}_* \right)} \,
	\mathcal{T}^\0 (\mathcal{E},\mathcal{S}) \, .
\end{equation}
We have absorbed into the effective partition function $Z''$
the leading term in the expansion~\eqref{eq_Laplace_exp_App}.
Again, $Z''$ is fixed by the normalization condition 
$\langle \id \rangle_\th = 1$.

We can approximate, and assess the scalings of, 
several components of the integral~\eqref{eq_k0_Therm_Int2_App}.
First, according to conventional thermodynamics,
$-\partial^2 \ent/\partial \mathcal{E}^2 \propto 1/(\text{heat capacity})$ is positive and $O(\Sites^{-1})$.
Therefore, we expect
$\left[-\nabla^2_\hess \ent \right]_*$ to be a positive-definite matrix whose elements are $O(\Sites^{-1}) \, .$
Therefore, the Gaussian factor in~\eqref{eq_k0_Therm_Int2_App} has a peak width of $O(\Sites^{1/2}) \, .$
Second, by evaluating the LHS of 
$\expval{H}_\th \equiv E \, ,$ using Eq.~\eqref{eq_k0_Therm_Int2_App},
we obtain $\mathcal{E}_* \approx E \, .$
Third, by evaluating the LHS of
$\langle \vec{S}^2 \rangle_\th 
= O(\Sites) \, ,$\footnote{
We evaluate $\langle \vec{S}^2 \rangle_\th$ by
replacing the $\mathcal{T}^\0 (\mathcal{E},\mathcal{S})$ in Eq.~\eqref{eq_k0_Therm_Int2_App} with $\mathcal{S}(\mathcal{S}+1) \, .$ We can understand the $O(\Sites)$ through the thermal state's sharing of scaling behaviors with the initial state:
$\langle \vec{S}^2 \rangle_\th 
\sim \langle \vec{S}^2 \rangle_0$
$= \expval{ S_x^2 + S_y^2 + S_z^2 }_0$
$= \var_0(S_x) + \var_0(S_y) + \var_0(S_z) + \langle S_z \rangle_0^2 \, .$
Each of the first three terms is $O(\Sites)$, by the variance conditions~\eqref{eq_Var_z}--\eqref{eq_Var_xy}. 
The final term vanishes because $M=0$ in this Supplementary Note.
Hence $\langle \vec{S}^2 \rangle_\th = O(\Sites) \, .$}
also using Eq.~\eqref{eq_k0_Therm_Int2_App},
we infer that 
$\mathcal{S}_* \in [0, \, O(\Sites^{1/2})] \, .$
Fourth, by the foregoing observations,
the Gaussian peaks far from the integration limits
$\mathcal{E} = E_\Min,  E_\Min$ and 
$\mathcal{S} = s_\Max \, .$
Therefore, we can extend these limits to $\pm \infty \, .$
Applying these conclusions to Eq.~\eqref{eq_k0_Therm_Int2_App} yields
\begin{equation}
    \label{eq_k0_Therm_Final_App}
	\expval{ T^\0_0 }_\th 
	\approx \frac{1}{ Z'' } 
	\int_{-\infty}^{\infty} d\mathcal{E} 
	\int_0^{\infty} d\mathcal{S} \
	e^{- \frac{1}{2} \left( \mathcal{E} - E, \, \mathcal{S} - \mathcal{S}_* \right)^T
	\left[-\nabla^2_\hess \ent \right]_*
	\left( \mathcal{E} - E, \, \mathcal{S} - \mathcal{S}_* \right)} \,
	\mathcal{T}^\0 (\mathcal{E},\mathcal{S}) \, .
\end{equation}

Whereas the exponential peaks sharply,
$\mathcal{T}^\0 (\mathcal{E},\mathcal{S})$
is smooth and varies slowly, by assumption
[see the text immediately above the non-Abelian ETH~\eqref{eq_nonAbelian_ETH}].
We therefore Taylor-expand 
$\mathcal{T}^\0 (\mathcal{E}, \mathcal{S})$ about $\mathcal{E} = E$ and $\mathcal{S} = 0$:\footnote{
Strictly speaking, one should Taylor-expand 
$\mathcal{T}^\0 (\mathcal{E}, \mathcal{S})$ about the maximum, $\mathcal{S} = \mathcal{S}_*$, rather than about $\mathcal{S} = 0$. However, $\mathcal{T}^\0$ is sufficiently smooth, and $\mathcal{S}_* \in [0, O(\Sites^{1/2})]$ is sufficiently close to $0$, that the two expansions yield identical results.}
\begin{align}
    \label{eq_T_k_critical}
	\mathcal{T}^\0 \left(\mathcal{E}, \mathcal{S} \right) 
	\approx \mathcal{T}^\0 \left(E, 0 \right) 
	+ O\! \left( \frac{\mathcal{E} - E}{\Sites} \right)
	+ O\! \left( \frac{\mathcal{S}}{\Sites} \right) + \dots
\end{align}
We argue for the nonzero $O(\mathcal{S}/\Sites)$ term's presence, for suitable Hamiltonians and suitable operators $\mathcal{T}^\0 \, ,$ in Suppl.~Note~\ref{app_Form_of_Tk}.

We substitute the Taylor expansion~\eqref{eq_T_k_critical} into Eq.~\eqref{eq_k0_Therm_Final_App}, 
then perform standard multivariate Gaussian integration.
The zeroth-order term in Eq.~\eqref{eq_T_k_critical} integrates to $\mathcal{T}^\0 (E, 0) \, ,$ by the integral's normalization.
The $O([\mathcal{E} - E]/\Sites)$ term integrates to zero, by the $\mathcal{E}$ integral's symmetry.
The $O(\mathcal{S}/\Sites)$ term does not integrate to zero similarly, because the $\mathcal{S}$ integral ends at 
$\mathcal{S} = 0 \, .$
However, the Gaussian has a width of
$O(\Sites^{1/2}) \, .$
Hence $\mathcal{S}$ averages to $O(\Sites^{1/2}) \, ,$
and the $O(\mathcal{S} / \Sites)$ term in Eq.~\eqref{eq_T_k_critical}
averages to $O(\Sites^{-1/2})$.
The higher-order terms average to $\leq O(\Sites^{-1})$, so
\begin{align}
    \expval{ T^\0_0 }_\th 
	\approx \mathcal{T}^\0 \left(E, 0 \right) 
	+ O ( \Sites^{-1/2} ) \, .
\end{align}

\end{document}